\documentclass[
 reprint,
superscriptaddress,
 amsmath,amssymb,
 aps,
prl,
longbibliography
]{revtex4-1}

\usepackage{graphicx}
\usepackage{dcolumn}
\usepackage{bm}
\usepackage{bbm}
\usepackage[utf8]{inputenc}
\usepackage{amsfonts}
\usepackage{amsthm}
\usepackage{tikz}
\usepackage{physics}
\usepackage{comment}
\usepackage{xcolor}
\usepackage[caption=false,justification=justified]{subfig}
\usepackage{blindtext}
\usepackage{pgfplots,pgfplotstable}
\usepgfplotslibrary{polar}
\usepgfplotslibrary{colormaps}
\pgfplotsset{compat=newest}
\usepackage{hyperref}
\usepackage{xr-hyper}
\usepackage{footmisc}
\usepackage{cancel}
\usepackage{cleveref}
\crefname{equation}{Eq.}{Eqs.}

\providecommand{\E}{\mathbb{E}}

\providecommand{\abs}[1]{\left\lvert#1\right\rvert}

\renewcommand{\d}{\mathrm{d}}
\newcommand{\e}{\mathrm{e}}
\newcommand{\ii}{\mathrm{i}}

\newcommand{\Var}[1]{\mathrm{Var}[#1]}
\newcommand{\PLH}{}

\begin{document}

\title{Estimation with ultimate quantum precision of the transverse displacement between two photons via two-photon interference sampling measurements}

\author{Danilo Triggiani}
\email{danilo.triggiani@port.ac.uk}
\affiliation{School of Mathematics and Physics, University of Portsmouth, Portsmouth PO1 3QL, UK}

\author{Vincenzo Tamma}
\email{vincenzo.tamma@port.ac.uk}
\affiliation{School of Mathematics and Physics, University of Portsmouth, Portsmouth PO1 3QL, UK}
\affiliation{Institute of Cosmology and Gravitation, University of Portsmouth, Portsmouth PO1 3FX, UK}

\date{\today}

\begin{abstract}
We present a quantum sensing scheme achieving the ultimate quantum sensitivity in the estimation of the transverse displacement between two photons interfering at a balanced beam splitter, based on transverse-momentum sampling measurements at the output.
This scheme can possibly lead to enhanced high-precision nanoscopic techniques, such as super-resolved single-molecule localization microscopy with quantum dots, by circumventing the requirements in standard direct imaging of cameras resolution at the diffraction limit, and of highly magnifying objectives.
Interestingly, we show that our interferometric technique achieves the ultimate spatial precision in nature irrespectively of the overlap of the two displaced photonic wavepackets, while its precision is only reduced of a constant factor for photons differing in any non-spatial degrees of freedom. 
This opens a new research paradigm based on the interface between spatially resolved quantum interference and quantum-enhanced spatial sensitivity.
\end{abstract}

\pacs{Valid PACS appear here}

\maketitle
The Hong-Ou-Mandel (HOM) effect~\cite{Hong1987,Shih1988} is an emblematic quantum phenomenon that proved to be instrumental for the development of novel quantum technologies~\cite{Bouchard2021}.
When two identical photons impinge on the two faces of a balanced beam splitter, and photodetectors are employed at both the outputs, no coincidence detection is recorded, as the two photons always 'bunch' in the same output channel.
This is caused by the quantum interference occurring between the two possible, but indistinguishable paths undertaken by the two identical photons through the beam splitter.
Instead, for non-identical photons, e.g., in their emission times, polarizations or central frequencies, the coincidence rate changes with the distinguishability between the photons at their detection~\cite{Hong1987, Shih1988, Bouchard2021}.
For this reason, HOM interferometry has been largely employed for high-precision measurements, e.g., of optical lengths and time delays~\cite{Hong1987, Lyons2018, Chen2019}, polarisations~\cite{Harnchaiwat2020, Sgobba2023}, and for quantum-enhanced imaging techniques, such as quantum optical coherence tomography~\cite{Abouraddy2002, Nasr2009}.
Furthermore, the analysis of the bounds on the precision achievable in an estimation protocol given by the Cramér-Rao bound (CRB)~\cite{Cramer1999, Rohatgi2000}, and of the ultimate precision achievable in nature through the quantum Cramér-Rao bound (QCRB)~\cite{Helstrom1969, Holevo2011}, has become a useful tool to determine the sensitivity of two-photon interferometry techniques for metrological applications~\cite{Lyons2018, Scott2020, Fabre2021, Johnson2023}.

Interestingly, it has been recently shown that inner-variables resolved two-photon interference, in which two delayed or frequency-shifted photons impinging on the two faces of the beam splitter are detected by frequency- or time-resolving detectors respectively, circumvents the requirement of large overlap in the photonic wavepackets typical of HOM interferometry~\cite{Legero2003, Legero2004, Tamma2015, Jin2015, Yepiz-Graciano2020, Triggiani2023, Wang2018, Hiemstra2020, Prakash2021}.
Indeed, this technique allows us to observe beating, i.e. oscillations, in probabilities of coincidence and bunching of the two photons, with a period that is inversely proportional to the difference in the colours or in the incidence times at the beam splitter, hence preserving its sensitivity also in the case of non-overlapping wavepackets~\cite{Triggiani2023}.

Two-photon interference has also been performed in the spatial domain, i.e. while varying transversal properties of the two photonic wavepackets, for example slightly rotating the momentum of the photons before they interfere~\cite{Ou1988, Kim2006}, manipulating the spatial overlap between the photons~\cite{Lee2006}, or simultaneously introducing a temporal delay to observe spatio-temporal coherence properties of the two-photon state~\cite{Lee2006, Devaux2020}. 
Spatial HOM interferometry so far has mostly been employed to assess the spatial coherence of highly entangled photons, e.g., produced by spontaneous parametric down-conversion.
However, to the best of our knowledge, the metrological potential of spatial two-photon interference for high precision imaging and sensing applications has not been investigated yet.
A study in this direction is made even more compelling, for an experimental and technological point of view, due to the recent development of high-precision nanoscopic techniques that already employ single-photon emitters and single-photon cameras~\cite{Hanne2015,Bruschini2019, Urban2021}.

\begin{figure}[t]
\includegraphics[width=.85\columnwidth]{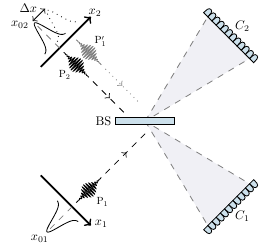}
\caption{Two single photons, centred in positions $x_{01}$ and $x_{02}$ in their respective transverse planes, impinge onto the two faces of a balanced beam splitter (BS), so that the displacement between the probe photon $\mathrm{P}_2$ and the symmetric image $\mathrm{P}_1'$ of the reference photon $\mathrm{P}_1$ is $\Delta x=x_{01}-x_{02}$.
The two photons are then detected by two cameras ($C_1$ and $C_2$) in the far-field regime, which resolve their transverse momenta $k\PLH$ and $k'$ for joint detections either in the same (bunching events) or opposite (coincidence events) output channels of the beam splitter.
}
\label{fig:Setup}
\end{figure}
This work introduces a quantum interference technique based on spatially resolved sampling measurements and investigates its properties from the metrological point of view. 
We show that resolving and sampling over the difference in transverse momenta of two photons detected after they interfere at a beam splitter, is an optimal metrological scheme for the estimation of the transverse separation of their wavepackets, achieving the ultimate precision given by the QCRB. 
This can be done by employing two cameras that spatially sample over all the possible two-photon interference events in the far field, resolving the difference in the transverse momenta of the two detected photons, and simultaneously recording whether the two photons ended up in the same or different output channels of the beam splitter (see \figurename~\ref{fig:Setup}).
This can therefore be seen as an interesting metrological application for spatial estimation of multiboson correlation sampling based on inner-variable sampling measurements~\cite{Laibacher2015, Tamma2016, Tamma2023}.
In particular, we show that, apart from a constant non-vanishing factor, the sensitivity of the proposed technique is retained for photons that also differ in any physical property other than their transverse position, e.g. their polarization or frequency, useful for localization and tracking applications with independent emitters.
Furthermore, the precision of this scheme can be, in principle, increased arbitrarily, for any fixed displacement between the photons, by employing photons with broader and broader transverse-momentum distributions.
Since our technique does not involve a direct detection of the single photons in the position domain, it removes the need of high-resolution cameras and highly magnifying objectives, typically employed in current localization microscopy techniques, which require to directly resolve sources at the diffraction limit~\cite{Ober2004, Lelek2021}.
Moreover, for almost identical photons, the resolving cameras can be replaced with bucket detectors that only count coincidences and bunching events, without affecting the sensitivity of the scheme.
Other than single molecule localization nanoscopy, this technique finds possible applications in astrophysical bodies localization, or for the measurement of any transverse displacement of a probe single-photon beam with respect to a reference beam, e.g. caused by refraction or by optical devices such as tuneable beam displacers~\cite{Serrano2015}.

\paragraph{Experimental setup.}
We consider pairs of quasi-monochromatic photons, with wavenumber $k_0$, produced by two independent sources impinging on the two faces of a balanced beam-splitter, and then detected by two cameras positioned at the beam-splitter outputs in the far-field regime at a distance $d$ from their sources.
For simplicity, we will only describe one transverse dimension of the setup, but the same analysis can be easily generalised to the two-dimensional case.
We suppose that the transverse position of the $S$-th photon is described by a wavepacket $\psi_S(x)=\psi(x-x_{0S})$ centred around the position $x_{0S}$, with $S=1,2$.
A practical example of such a setup is shown in the schematic representation of \figurename~\ref{fig:Setup}.
We can thus write the two-photon state as
\begin{equation}
\ket{\Psi} = \int\d x_1\ \psi_1(x_1)\hat{a}_1^\dag(x_1) \ket{0}_1 \otimes \int\d x_2\ \psi_2(x_2)\hat{b}_2^\dag(x_2)\ket{0}_2,
\label{eq:InitState}
\end{equation}
where $\hat{a}^\dag_1(x_1)$ and $\hat{b}^\dag_2(x_2)$ are the bosonic creation operators associated with the first and second input mode of the beam-splitter at transverse positions $x_1$ and $x_2$ respectively, satisfying the commutation relation $\comm{\hat{a}_S(x)}{\hat{b}^\dag_{S'}(x')} = \sqrt{\nu}\delta_{SS'}\delta(x-x')$, with $S,S'=1,2$, where $\delta_{SS'}$ and $\delta(x-x')$ denote the Kronecker and Dirac delta respectively, while $0\leqslant\nu\leqslant 1$ represents a degree of indistinguishability between the photons in any physical property other than their transverse positions (e.g. different polarisations), with $\nu=1$ denoting the maximum degree of indistinguishability.
The displacement $\Delta x=x_{01}-x_{02}$ that we aim to measure can be caused by refraction or any other beam-displacing optical device~\cite{Serrano2015}, or it can represent the position of a single-photon source that we want to localise, e.g. a probe quantum dot attached to a molecule~\cite{Hanne2015, Bruschini2019, Urban2021}, with respect to an identical reference photon emitted at a known position.

\begin{figure}
\includegraphics[width=.95\columnwidth]{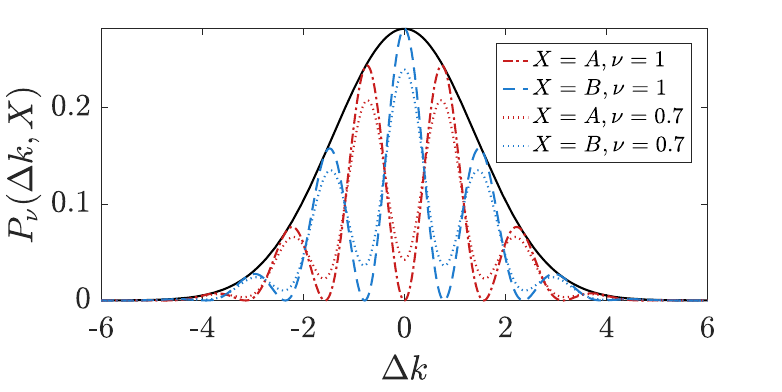}
\caption{Plots of the probability $P_\nu(\Delta k,X)$ in Eq.~\eqref{eq:Prob} for Gaussian transverse-momentum distributions $\abs{\varphi(k)}^2$, yielding a Gaussian envelope $C(\Delta x)$ (black solid line), for same-camera ($X=B$) detection events, and different cameras ($X=A$) detection events.
The variance of the distribution $\abs{\varphi(k)}^2$ is set to $\sigma_k^2=1$, which fixes a natural scale for $\Delta k$ and for the separation $\Delta x=4/\sigma_k$.
}
\label{fig:Probabilities}
\end{figure}

After impinging on the beam-splitter, each pair of photons is randomly detected at the pixels in positions $y\PLH$ and $y'$ either of a single or distinct cameras. 
Since the detection occurs in the far-field regime, it corresponds to resolving the transverse momenta $k\PLH = y\PLH k_0/d$ and $k' = y' k_0/d$ of the two photons, i.e. the conjugate variables to the photon transverse positions.
We show that this results in the observation of quantum beats in the difference $\Delta k=k\PLH-k'$ (see \figurename~\ref{fig:Probabilities}) with periodicity inversely proportional to $\Delta x = x_{01}-x_{02}$ in the joint probabilities
\begin{equation}
P_\nu(\Delta k,X) = \frac{1}{2}C(\Delta k)\left(1+\alpha(X)\nu\cos(\Delta k\Delta x)\right),
\label{eq:Prob}
\end{equation}
of the two photons detected by different cameras ($X=A$), or the same camera ($X=B$), where $\alpha(A) = -1$ and $\alpha(B)=+1$, and $C(\Delta k)$ is the beats envelope, whose shape is known and depends on the transverse-momentum distribution of the photons, i.e. the modulo squared $\abs{\varphi(k)}^2$ of the Fourier transform  of $\psi(x)$, e.g. $C(\Delta k)=\exp(-\Delta k^2/4\sigma_k^2)/\sqrt{4\pi\sigma_k^2}$ for Gaussian wavepackets~\cite{Suppl}\nocite{Glauber1963}.

The estimation of the displacement $\Delta x$ is carried out by performing a given number $N$ of sampling measurements of the values $(\Delta k, X)$ with pixel size $\delta y$,  corresponding to a precision $\delta k=\delta y k_0/d$ in the transverse momenta, small enough to resolve the transverse-momentum distributions of the two photons and the beating oscillations with period $2 \pi/ \Delta x$  in Eq.\eqref{eq:Prob}, i.e. 
\begin{equation}
\delta y\ll \frac{d}{k_0}\sigma_k,\quad \delta y\ll \frac{d}{k_0}\frac{1}{\Delta x},
\label{eq:Resolution}
\end{equation}
where $\sigma_k^2$ is the variance of the transverse-momentum distribution of the photons.
The $N$ detected sampling outcomes $(\Delta k_i, X_i)$ with $i =1,\dots, N$  are than employed in a standard maximum-likelihood estimation procedure of the parameter $\Delta x$, specialized to the probability distribution in Eq.~\eqref{eq:Prob}~\cite{Suppl}.

Due to the Fourier uncertainty principle $\sigma_k\geqslant 1/2\sigma_x$, with $\sigma_x$ given by the standard deviation of $\abs{\psi(x)}^2$ and which, for diffraction limited optics, can be approximated by $\sigma_x\simeq 2\pi/2.8k_0\simeq 2.2/k_0$, it is possible to express the first resolution requirement in Eq.~\eqref{eq:Resolution} as $\delta y\ll 0.22 d$, easily satisfied with current cameras, independently of the color of the photons.
Furthermore, since this technique requires cameras only able to resolve oscillations of period $\propto 1/\Delta x$, it circumvents the need to employ objectives with high magnifying factors to resolve directly the relative position $\Delta x$ of the single-photon emitters at the diffraction limit, typically required in imaging techniques such as single-molecule localization microscopy~\cite{Ober2004, Lelek2021}.

\paragraph{Ultimate quantum sensitivity.}

\begin{figure}
\centering
\includegraphics[width=\columnwidth]{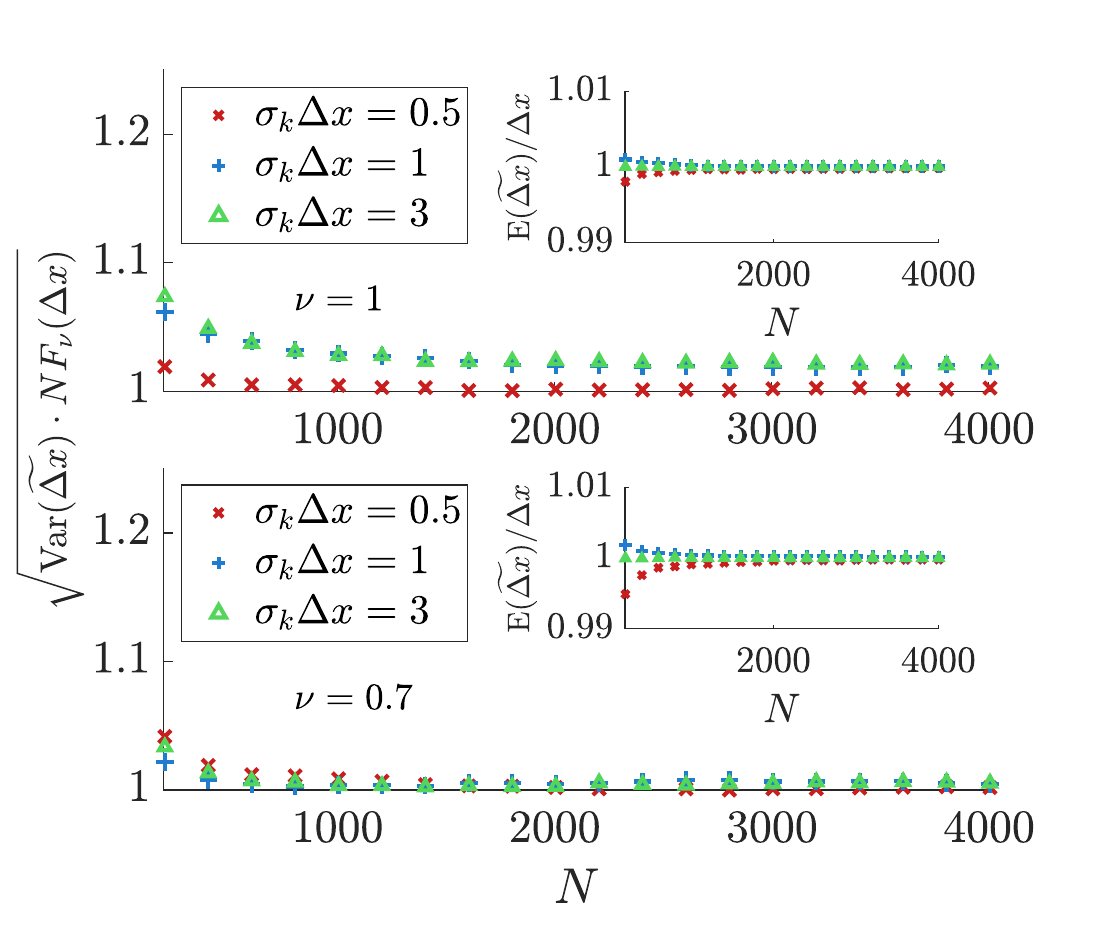}
\caption{Numerical simulations for Gaussian photonic wavepackets and for different values of $\sigma_k\Delta x$ of the variance $\mathrm{Var}(\widetilde{\Delta x})$ of the maximum likelihood estimator $\widetilde{\Delta x}$, which saturates the CRB in Eq.~\eqref{eq:QCRB} for the FI $F_\nu(\Delta x)$ in Eq.~\eqref{eq:FINu} at the increasing of the number $N$ of sampling measurement iterations of the estimation.
In the inset, we plot the ratio between the expected value $\mathrm{E}(\widetilde{\Delta x})$ of the maximum-likelihood estimator and the true value $\Delta x$.
}
\label{fig:Saturation}
\end{figure}

Any unbiased estimator $\widetilde{\Delta x}$ associated with   the estimation of the separation $\Delta x$ between the photon pair in Eq.~\eqref{eq:InitState}, will have a variance $\Var{\widetilde{\Delta x}}$ bounded from below by the CRB associated with the described measurement~\cite{Cramer1999, Rohatgi2000}, and ultimately by the QCRB, which sets the ultimate quantum limit of the achievable precision independently of the measurements~\cite{Helstrom1969, Paris2009, Holevo2011, Liu2020}, i.e.
\begin{equation}
\Var{\widetilde{\Delta x}}\geqslant \frac{1}{N F_\nu(\Delta x)}\geqslant \frac{1}{N H(\Delta x)},
\label{eq:QCRB}
\end{equation}
where $H(\Delta x)$ is the quantum Fisher information (QFI),  $F_\nu(\Delta x)$ the Fisher information (FI), and $N$ is the number of independent sampling measurements~\cite{Cramer1999,Helstrom1969}.
In particular, we demonstrate from Eq.~\eqref{eq:Prob} that in our scheme, in case of photons differing only in their transverse positions,
\begin{equation}
F_{\nu=1}(\Delta x) = H(\Delta x) \equiv H \equiv 2 \sigma^2_k,
\label{eq:QFI}
\end{equation}
i.e. the FI saturates the QFI~\cite{Suppl}.
We show in \figurename~\ref{fig:Saturation} that the QCRB in Eq.~\eqref{eq:QCRB} is already approximately saturated for $N$ of the order of $1000$ by maximising the measured  likelihood function, calculated in the Supplemental Material, i.e.
\begin{equation}
\Var{\widetilde{\Delta x}}\simeq \frac{1}{N H}= \frac{1}{2N\sigma_k^2}.
\label{eq:QCRBN}
\end{equation} 

Remarkably, the ultimate precision in Eq.~\eqref{eq:QCRBN} achieved with our technique is independent of the separation $\Delta x$ to be estimated, and can be, in principle, arbitrarily improved by increasing the transverse-momentum variance $\sigma_k^2$ of the photons.
For example, assuming Gaussian distributions $\abs{\psi(x)}^2$ with a diffraction limited width $\sigma_x\simeq\! 100$nm~\cite{Lelek2021}, the bound on the sensitivity in Eq.~\eqref{eq:QCRBN} amounts to $\sqrt{1/(2N\sigma_k^2)}\simeq 2.6$nm employing only $N\simeq3000$ pairs of photons.
This would not be possible in direct imaging without the use of magnifying objectives, where even by employing a camera with $\sim\!10\mathrm{\mu}$m pixel size, x$100$ magnifying lenses would be required to reach the diffraction limit. Localising single-photon wave packets with smaller and smaller diffraction limited widths becomes increasingly prohibitive for direct imaging techniques, due to the need of lenses with higher and higher magnification factors and the detrimental effects of misalignments and aberrations~\cite{Khater2020}, while, in our technique, not only it is feasible, but also allows according to Eq.~\eqref{eq:QCRBN} higher and higher precisions, useful for example for studies of structures and functions at the sub-cellular level~\cite{Sahl2017}.  
Furthermore, already for $N\simeq3000$ sampling iterations, we achieve a 97\% saturation of the CRB, and a relative bias smaller than  $0.1\%$, without the need of fully retrieving the output probability distribution with a larger number of measurements (see \figurename~\ref{fig:Saturation}).

Interestingly, this interferometric technique is completely unaffected by the lack of overlap between the spatial wavepackets of the two photons.
Indeed, performing transverse-momentum resolving detections, i.e., resolving the variables in the conjugate domain to the position, renders the detectors ``blind'' to the position of emission of the two photons, and thus it enables the observation of two-photon quantum interference.
This feature is in stark contrast with standard non-resolving two-photon interference, where the distinguishability of photons with spatially non-overlapping wavepackets is not erased at the detectors, hindering the precision of the estimation.
Nevertheless, for photons with mostly overlapping spatial wavepackets, i.e. for $\sigma_k\Delta x\ll1$, we show in the Supplemental Material that it is possible to saturate the QFI without resolving their transverse momenta and therefore by using simple bucket detectors.

\begin{figure}
\includegraphics[width=\columnwidth]{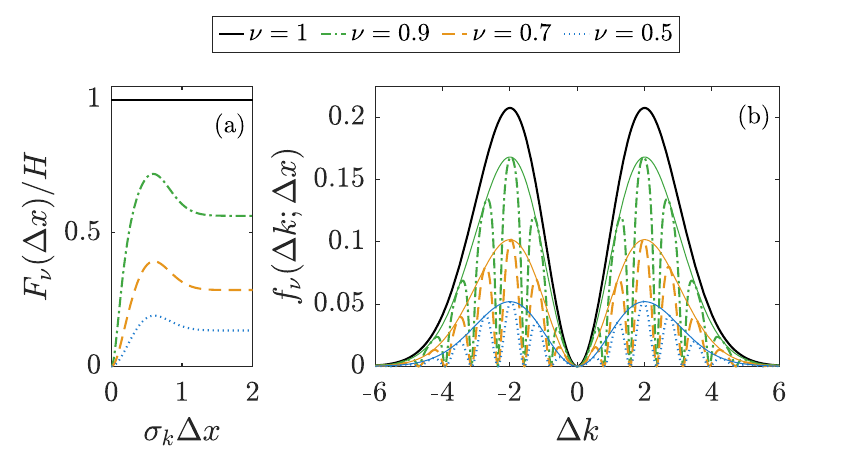}
\caption{Plots for Gaussian transverse-momentum distributions $\abs{\varphi(k)}^2$ and different values of $\nu$ of: (a) $F_\nu(\Delta x)$ normalised to $H$ in Eq.~\eqref{eq:FINu}, with Eq.~\eqref{eq:FIgtr} holding independently of the value of $\Delta x$, for $\sigma_k\Delta x\gtrsim 0.5$; (b) Contributions $f_\nu(\Delta k;\Delta x)$ in Eq.~\eqref{eq:FIContr} to the FI depending on the outcome value $\Delta k$ for $\Delta x=4/\sigma_k$, with $\sigma_k=1$ which fixes a natural scale for $\Delta k$.
All the information on $\Delta x$ can be obtained by sampling measurements in the transverse momenta difference $\Delta k$ within its envelopes independently of the value of $\Delta x$ (thin solid lines).}
\label{fig:FIContr}
\end{figure}

\paragraph{FI for partially distinguishable detected photons.}
Our technique remains effective also when introducing partial distinguishability $\eta<1$ in the two-photon detection.
Indeed, in this more general scenario, we show, by using Eq.~\eqref{eq:Prob}, that the FI, normalised to the QFI $H$ in Eq.~\eqref{eq:QFI}, reads~\cite{Suppl}
\begin{equation}
\frac{F_\nu(\Delta x)}{H} = \int \dd\Delta k\ f_\nu(\Delta k;\Delta x),
\label{eq:FINu}
\end{equation}
where
\begin{equation}
f_\nu(\Delta k;\Delta x)=C(\Delta k)\frac{\Delta k^2}{2\sigma_k^2}\frac{\nu^2\sin^2(\Delta k\Delta x)}{1-\nu^2\cos^2(\Delta k\Delta x)}.
\label{eq:FIContr}
\end{equation}
As evident in \figurename~\ref{fig:Saturation}, also for $\nu<1$, the CRB is approximately saturated already with $N$ of the order of $1000$, i.e. $\Var{\widetilde{\Delta x}}\simeq 1/NF_\nu(\Delta x)$.
Importantly, from the plot of $F_\nu(\Delta x)$ in \figurename~\ref{fig:FIContr}, it is evident that, for $\sigma_k\Delta x\gtrsim 0.5$, condition that can always be guaranteed by calibrating the position of the reference photon, for Gaussian wavepackets 
\begin{equation}
F_\nu(\Delta x)\gtrsim {(1-\sqrt{1-\nu^2})}H,
\label{eq:FIgtr}
\end{equation} 
independently of the value of $\Delta x$ (e.g. with $1-\sqrt{1-\nu^2}\simeq 0.6$ for $\nu=0.9$), where ${(1-\sqrt{1-\nu^2})}H$ is the asymptotic value for $\sigma_k\Delta x\gg 1$~\cite{Suppl}.

\paragraph{Contribution from the sampled transverse momenta.}
Each term $f_\nu(\Delta k;\Delta x)$ in Eq.~\eqref{eq:FIContr} represents, apart from a factor $2\sigma_k^2$, the contribution to $F_\nu(\Delta x)$ in Eq.~\eqref{eq:FINu} yielded by both outcomes $(\Delta k,A)$ and $(\Delta k,B)$ of a single two-photon detection.
The expression of $f_\nu(\Delta k;\Delta x)$ in Eq.~\eqref{eq:FIContr} allows us to understand which outcome values of $\Delta k$ yield more information on the separation $\Delta x$, and conversely which ones can be discarded with a negligible loss of precision.
Indeed, we notice from \figurename~\ref{fig:FIContr}b that  $f_\nu(\Delta k;\Delta x)$, as a function of $\Delta k$, is concentrated within its envelope $\nu^2 C(\Delta k)\Delta k^2/2\sigma_k^2$, associated with the distributions of the photons in the transverse momenta, independently of $\Delta x$. 
Therefore, the cameras do not need to be adjusted according to the unknown separation $\Delta x$ between the
two photons. Furthermore,  if the tails of the distribution of the photons have a negligible contribution (e.g. for Gaussian wavepackets), one can integrate the $f_\nu(\Delta k;\Delta x)$ in Eq.~\eqref{eq:FIContr} within a few units of $\sigma_k$, hence with a reduced cameras sensing range.

\paragraph{Contribution from single or joint camera detections.}
\begin{figure}
\includegraphics[width=\columnwidth]{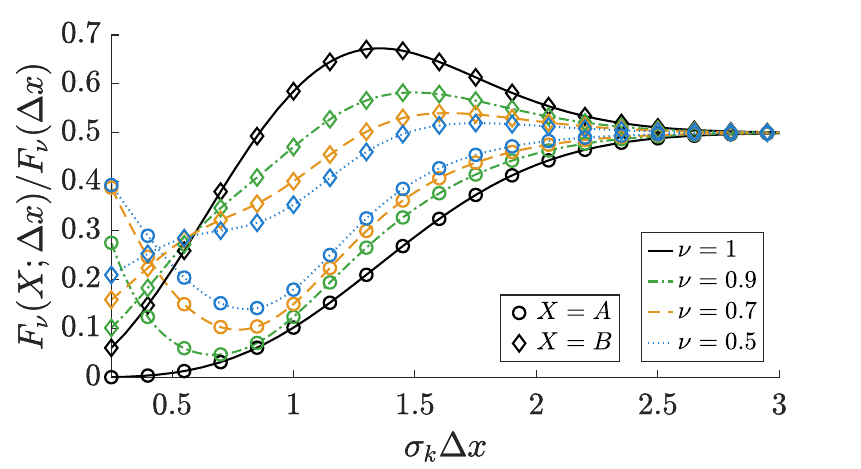}
\caption{Plot of the ratios $F_\nu(X;\Delta x)/F_\nu(\Delta x)$ between the FI associated with the observation of only two-camera ($X=B$), or only single-camera events ($X=A$), and the FI in Eq.~\eqref{eq:FINu} of our original scheme, for different values of $\nu$. Observing only single-camera events yields a better precision than double-camera events for $\sigma_k\Delta x\gtrsim 0.5$.}
\label{fig:FPartial}
\end{figure}

In \figurename~\ref{fig:FPartial} we show, for different values of $\nu$, the FI $F_\nu(X;\Delta x)$ associated with the observation of only two cameras ($X=A$) events or single camera ($X=B$) events, according to  the probability distributions $P_\nu(\Delta k,X)$ in Eq.~\eqref{eq:Prob}, normalized over the overall probability $\int \dd \Delta k\ P_\nu(\Delta k,X)$~\cite{Suppl}. 
For $\sigma_k\Delta x\gtrsim 0.5$ and Gaussian photonic wavepackets, we see from \figurename~\ref{fig:FPartial} that observing only one-camera events ($X=B$) is more informative than observing only two-camera events ($X=A$), since $F_\nu(B;\Delta x)\geqslant F_\nu(A;\Delta x)$, and $F_\nu (B; \Delta x)$ is only reduced, compared to $F_\nu(\Delta x)$, by a non-vanishing factor $R_\nu=F_\nu (B; \Delta x)/F_\nu (\Delta x)$  (e.g. $R_\nu \simeq 0.6$ for $\nu=0.9$ and $R_\nu \simeq 0.7$ for $\nu=1$, for $\sigma_k \Delta x \simeq 1.5$).
Therefore, by using Eq.~\eqref{eq:FIgtr}, we deduce that our scheme can also be employed with only one camera, achieving a FI $\gtrsim H R_\nu(1-\sqrt{1-\nu^2})/2$, where the factor $1/2$ provides for the fact that we are observing only one output channel of the beam splitter.

\paragraph{Conclusions.}

We presented a spatial quantum interference technique allowing the estimation of the transverse separation of two single-photons beams at the ultimate precision allowed by the QCRB, based on transverse-momentum resolved sampling measurements. In the regime of almost identical photons, the two cameras can be replaced with non-resolving bucket detectors without any loss of sensitivity.
Furthermore, apart from a non-vanishing factor, the same quantum sensitivity is retained also for photons partially differing in any physical parameters other than their transverse position, and it can be also achieved by employing only one camera at a single output of the beam splitter. 
Finally, such an ultimate quantum sensitivity can be approximately obtained for a limited number of iterations of the experiment, and be, in principle, arbitrarily increased by employing photons with more broadly distributed transverse momenta. This means that such a precision is only limited by the highest value of the transverse-momentum distribution variance experimentally feasible with current technologies.
Moreover, by using indirect sampling measurements which resolve the photonic transverse momenta instead of the positions directly, this technique removes the requirement of cameras resolution at the diffraction limit typical of direct imaging techniques, as well as highly magnifying objectives.

These results shed new light on the metrological power of two-photon spatial interference and can pave the way to new high-precision sensing techniques.
This work has the potential to enhance super-resolution imaging techniques that already employ single-photon sources as probes for the localization and tracking of biological samples, such as single-molecule localization microscopy with quantum dots~\cite{Hanne2015, Bruschini2019, Urban2021}.
Other possible applications can be found in the development of quantum sensing techniques for high-precision refractometry and astrophysical bodies localization, as well as high-precision multiparameter sensing schemes, including 3D quantum localization methods, and simultaneous estimation of the colour and position of single-photon emitters, by combining our technique with multiphoton interference techniques in the frequency-time domain~\cite{Lyons2018,Triggiani2023}.

\section*{Acknowledgements}

VT acknowledges support from the Air Force Office of Scientific Research under award number FA8655-23-1-7046.

\bibliography{references}

\begin{thebibliography}{44}%
\makeatletter
\providecommand \@ifxundefined [1]{%
 \@ifx{#1\undefined}
}%
\providecommand \@ifnum [1]{%
 \ifnum #1\expandafter \@firstoftwo
 \else \expandafter \@secondoftwo
 \fi
}%
\providecommand \@ifx [1]{%
 \ifx #1\expandafter \@firstoftwo
 \else \expandafter \@secondoftwo
 \fi
}%
\providecommand \natexlab [1]{#1}%
\providecommand \enquote  [1]{``#1''}%
\providecommand \bibnamefont  [1]{#1}%
\providecommand \bibfnamefont [1]{#1}%
\providecommand \citenamefont [1]{#1}%
\providecommand \href@noop [0]{\@secondoftwo}%
\providecommand \href [0]{\begingroup \@sanitize@url \@href}%
\providecommand \@href[1]{\@@startlink{#1}\@@href}%
\providecommand \@@href[1]{\endgroup#1\@@endlink}%
\providecommand \@sanitize@url [0]{\catcode `\\12\catcode `\$12\catcode
  `\&12\catcode `\#12\catcode `\^12\catcode `\_12\catcode `\%12\relax}%
\providecommand \@@startlink[1]{}%
\providecommand \@@endlink[0]{}%
\providecommand \url  [0]{\begingroup\@sanitize@url \@url }%
\providecommand \@url [1]{\endgroup\@href {#1}{\urlprefix }}%
\providecommand \urlprefix  [0]{URL }%
\providecommand \Eprint [0]{\href }%
\providecommand \doibase [0]{http://dx.doi.org/}%
\providecommand \selectlanguage [0]{\@gobble}%
\providecommand \bibinfo  [0]{\@secondoftwo}%
\providecommand \bibfield  [0]{\@secondoftwo}%
\providecommand \translation [1]{[#1]}%
\providecommand \BibitemOpen [0]{}%
\providecommand \bibitemStop [0]{}%
\providecommand \bibitemNoStop [0]{.\EOS\space}%
\providecommand \EOS [0]{\spacefactor3000\relax}%
\providecommand \BibitemShut  [1]{\csname bibitem#1\endcsname}%
\let\auto@bib@innerbib\@empty
\bibitem [{\citenamefont {Hong}\ \emph {et~al.}(1987)\citenamefont {Hong},
  \citenamefont {Ou},\ and\ \citenamefont {Mandel}}]{Hong1987}%
  \BibitemOpen
  \bibfield  {author} {\bibinfo {author} {\bibfnamefont {C.~K.}\ \bibnamefont
  {Hong}}, \bibinfo {author} {\bibfnamefont {Z.~Y.}\ \bibnamefont {Ou}}, \ and\
  \bibinfo {author} {\bibfnamefont {L.}~\bibnamefont {Mandel}},\ }\bibfield
  {title} {\enquote {\bibinfo {title} {Measurement of subpicosecond time
  intervals between two photons by interference},}\ }\href {\doibase
  10.1103/PhysRevLett.59.2044} {\bibfield  {journal} {\bibinfo  {journal}
  {Phys. Rev. Lett.}\ }\textbf {\bibinfo {volume} {59}},\ \bibinfo {pages}
  {2044--2046} (\bibinfo {year} {1987})}\BibitemShut {NoStop}%
\bibitem [{\citenamefont {Shih}\ and\ \citenamefont {Alley}(1988)}]{Shih1988}%
  \BibitemOpen
  \bibfield  {author} {\bibinfo {author} {\bibfnamefont {Y.~H.}\ \bibnamefont
  {Shih}}\ and\ \bibinfo {author} {\bibfnamefont {C.~O.}\ \bibnamefont
  {Alley}},\ }\bibfield  {title} {\enquote {\bibinfo {title} {New type of
  einstein-podolsky-rosen-bohm experiment using pairs of light quanta produced
  by optical parametric down conversion},}\ }\href {\doibase
  10.1103/PhysRevLett.61.2921} {\bibfield  {journal} {\bibinfo  {journal}
  {Phys. Rev. Lett.}\ }\textbf {\bibinfo {volume} {61}},\ \bibinfo {pages}
  {2921--2924} (\bibinfo {year} {1988})}\BibitemShut {NoStop}%
\bibitem [{\citenamefont {Bouchard}\ \emph {et~al.}(2021)\citenamefont
  {Bouchard}, \citenamefont {Sit}, \citenamefont {Zhang}, \citenamefont
  {Fickler}, \citenamefont {Miatto}, \citenamefont {Yao}, \citenamefont
  {Sciarrino},\ and\ \citenamefont {Karimi}}]{Bouchard2021}%
  \BibitemOpen
  \bibfield  {author} {\bibinfo {author} {\bibfnamefont {Fr{\'{e}}d{\'{e}}ric}\
  \bibnamefont {Bouchard}}, \bibinfo {author} {\bibfnamefont {Alicia}\
  \bibnamefont {Sit}}, \bibinfo {author} {\bibfnamefont {Yingwen}\ \bibnamefont
  {Zhang}}, \bibinfo {author} {\bibfnamefont {Robert}\ \bibnamefont {Fickler}},
  \bibinfo {author} {\bibfnamefont {Filippo~M}\ \bibnamefont {Miatto}},
  \bibinfo {author} {\bibfnamefont {Yuan}\ \bibnamefont {Yao}}, \bibinfo
  {author} {\bibfnamefont {Fabio}\ \bibnamefont {Sciarrino}}, \ and\ \bibinfo
  {author} {\bibfnamefont {Ebrahim}\ \bibnamefont {Karimi}},\ }\bibfield
  {title} {\enquote {\bibinfo {title} {Two-photon interference: the
  hong-ou-mandel effect},}\ }\href {\doibase 10.1088/1361-6633/abcd7a}
  {\bibfield  {journal} {\bibinfo  {journal} {Reports on Progress in Physics}\
  }\textbf {\bibinfo {volume} {84}},\ \bibinfo {pages} {012402} (\bibinfo
  {year} {2021})}\BibitemShut {NoStop}%
\bibitem [{\citenamefont {Lyons}\ \emph {et~al.}(2018)\citenamefont {Lyons},
  \citenamefont {Knee}, \citenamefont {Bolduc}, \citenamefont {Roger},
  \citenamefont {Leach}, \citenamefont {Gauger},\ and\ \citenamefont
  {Faccio}}]{Lyons2018}%
  \BibitemOpen
  \bibfield  {author} {\bibinfo {author} {\bibfnamefont {Ashley}\ \bibnamefont
  {Lyons}}, \bibinfo {author} {\bibfnamefont {George~C.}\ \bibnamefont {Knee}},
  \bibinfo {author} {\bibfnamefont {Eliot}\ \bibnamefont {Bolduc}}, \bibinfo
  {author} {\bibfnamefont {Thomas}\ \bibnamefont {Roger}}, \bibinfo {author}
  {\bibfnamefont {Jonathan}\ \bibnamefont {Leach}}, \bibinfo {author}
  {\bibfnamefont {Erik~M.}\ \bibnamefont {Gauger}}, \ and\ \bibinfo {author}
  {\bibfnamefont {Daniele}\ \bibnamefont {Faccio}},\ }\bibfield  {title}
  {\enquote {\bibinfo {title} {Attosecond-resolution hong-ou-mandel
  interferometry},}\ }\href
  {https://advances.sciencemag.org/content/4/5/eaap9416} {\bibfield  {journal}
  {\bibinfo  {journal} {Science Advances}\ }\textbf {\bibinfo {volume} {4}},\
  \bibinfo {pages} {5} (\bibinfo {year} {2018})}\BibitemShut {NoStop}%
\bibitem [{\citenamefont {Chen}\ \emph {et~al.}(2019)\citenamefont {Chen},
  \citenamefont {Fink}, \citenamefont {Steinlechner}, \citenamefont {Torres},\
  and\ \citenamefont {Ursin}}]{Chen2019}%
  \BibitemOpen
  \bibfield  {author} {\bibinfo {author} {\bibfnamefont {Yuanyuan}\
  \bibnamefont {Chen}}, \bibinfo {author} {\bibfnamefont {Matthias}\
  \bibnamefont {Fink}}, \bibinfo {author} {\bibfnamefont {Fabian}\ \bibnamefont
  {Steinlechner}}, \bibinfo {author} {\bibfnamefont {Juan~P.}\ \bibnamefont
  {Torres}}, \ and\ \bibinfo {author} {\bibfnamefont {Rupert}\ \bibnamefont
  {Ursin}},\ }\bibfield  {title} {\enquote {\bibinfo {title} {Hong-ou-mandel
  interferometry on a biphoton beat note},}\ }\href {\doibase
  10.1038/s41534-019-0161-z} {\bibfield  {journal} {\bibinfo  {journal} {npj
  Quantum Information}\ }\textbf {\bibinfo {volume} {5}},\ \bibinfo {pages}
  {43} (\bibinfo {year} {2019})}\BibitemShut {NoStop}%
\bibitem [{\citenamefont {Harnchaiwat}\ \emph {et~al.}(2020)\citenamefont
  {Harnchaiwat}, \citenamefont {Zhu}, \citenamefont {Westerberg}, \citenamefont
  {Gauger},\ and\ \citenamefont {Leach}}]{Harnchaiwat2020}%
  \BibitemOpen
  \bibfield  {author} {\bibinfo {author} {\bibfnamefont {Natapon}\ \bibnamefont
  {Harnchaiwat}}, \bibinfo {author} {\bibfnamefont {Feng}\ \bibnamefont {Zhu}},
  \bibinfo {author} {\bibfnamefont {Niclas}\ \bibnamefont {Westerberg}},
  \bibinfo {author} {\bibfnamefont {Erik}\ \bibnamefont {Gauger}}, \ and\
  \bibinfo {author} {\bibfnamefont {Jonathan}\ \bibnamefont {Leach}},\
  }\bibfield  {title} {\enquote {\bibinfo {title} {Tracking the polarisation
  state of light via hong-ou-mandel interferometry},}\ }\href {\doibase
  10.1364/OE.382622} {\bibfield  {journal} {\bibinfo  {journal} {Opt. Express}\
  }\textbf {\bibinfo {volume} {28}},\ \bibinfo {pages} {2210--2220} (\bibinfo
  {year} {2020})}\BibitemShut {NoStop}%
\bibitem [{\citenamefont {Sgobba}\ \emph {et~al.}(2023)\citenamefont {Sgobba},
  \citenamefont {Pallotti}, \citenamefont {Elefante}, \citenamefont
  {Dello~Russo}, \citenamefont {Dequal}, \citenamefont {Siciliani~de Cumis},\
  and\ \citenamefont {Santamaria~Amato}}]{Sgobba2023}%
  \BibitemOpen
  \bibfield  {author} {\bibinfo {author} {\bibfnamefont {Fabrizio}\
  \bibnamefont {Sgobba}}, \bibinfo {author} {\bibfnamefont {Deborah~Katia}\
  \bibnamefont {Pallotti}}, \bibinfo {author} {\bibfnamefont {Arianna}\
  \bibnamefont {Elefante}}, \bibinfo {author} {\bibfnamefont {Stefano}\
  \bibnamefont {Dello~Russo}}, \bibinfo {author} {\bibfnamefont {Daniele}\
  \bibnamefont {Dequal}}, \bibinfo {author} {\bibfnamefont {Mario}\
  \bibnamefont {Siciliani~de Cumis}}, \ and\ \bibinfo {author} {\bibfnamefont
  {Luigi}\ \bibnamefont {Santamaria~Amato}},\ }\bibfield  {title} {\enquote
  {\bibinfo {title} {Optimal measurement of telecom wavelength single photon
  polarisation via hong-ou-mandel interferometry},}\ }\href {\doibase
  10.3390/photonics10010072} {\bibfield  {journal} {\bibinfo  {journal}
  {Photonics}\ }\textbf {\bibinfo {volume} {10}} (\bibinfo {year} {2023}),\
  10.3390/photonics10010072}\BibitemShut {NoStop}%
\bibitem [{\citenamefont {Abouraddy}\ \emph {et~al.}(2002)\citenamefont
  {Abouraddy}, \citenamefont {Nasr}, \citenamefont {Saleh}, \citenamefont
  {Sergienko},\ and\ \citenamefont {Teich}}]{Abouraddy2002}%
  \BibitemOpen
  \bibfield  {author} {\bibinfo {author} {\bibfnamefont {Ayman~F.}\
  \bibnamefont {Abouraddy}}, \bibinfo {author} {\bibfnamefont {Magued~B.}\
  \bibnamefont {Nasr}}, \bibinfo {author} {\bibfnamefont {Bahaa E.~A.}\
  \bibnamefont {Saleh}}, \bibinfo {author} {\bibfnamefont {Alexander~V.}\
  \bibnamefont {Sergienko}}, \ and\ \bibinfo {author} {\bibfnamefont
  {Malvin~C.}\ \bibnamefont {Teich}},\ }\bibfield  {title} {\enquote {\bibinfo
  {title} {Quantum-optical coherence tomography with dispersion
  cancellation},}\ }\href {\doibase 10.1103/PhysRevA.65.053817} {\bibfield
  {journal} {\bibinfo  {journal} {Phys. Rev. A}\ }\textbf {\bibinfo {volume}
  {65}},\ \bibinfo {pages} {053817} (\bibinfo {year} {2002})}\BibitemShut
  {NoStop}%
\bibitem [{\citenamefont {Nasr}\ \emph {et~al.}(2009)\citenamefont {Nasr},
  \citenamefont {Goode}, \citenamefont {Nguyen}, \citenamefont {Rong},
  \citenamefont {Yang}, \citenamefont {Reinhard}, \citenamefont {Saleh},\ and\
  \citenamefont {Teich}}]{Nasr2009}%
  \BibitemOpen
  \bibfield  {author} {\bibinfo {author} {\bibfnamefont {Magued~B.}\
  \bibnamefont {Nasr}}, \bibinfo {author} {\bibfnamefont {Darryl~P.}\
  \bibnamefont {Goode}}, \bibinfo {author} {\bibfnamefont {Nam}\ \bibnamefont
  {Nguyen}}, \bibinfo {author} {\bibfnamefont {Guoxin}\ \bibnamefont {Rong}},
  \bibinfo {author} {\bibfnamefont {Linglu}\ \bibnamefont {Yang}}, \bibinfo
  {author} {\bibfnamefont {Björn~M.}\ \bibnamefont {Reinhard}}, \bibinfo
  {author} {\bibfnamefont {Bahaa~E.A.}\ \bibnamefont {Saleh}}, \ and\ \bibinfo
  {author} {\bibfnamefont {Malvin~C.}\ \bibnamefont {Teich}},\ }\bibfield
  {title} {\enquote {\bibinfo {title} {Quantum optical coherence tomography of
  a biological sample},}\ }\href {\doibase
  https://doi.org/10.1016/j.optcom.2008.11.061} {\bibfield  {journal} {\bibinfo
   {journal} {Optics Communications}\ }\textbf {\bibinfo {volume} {282}},\
  \bibinfo {pages} {1154--1159} (\bibinfo {year} {2009})}\BibitemShut {NoStop}%
\bibitem [{\citenamefont {Cram{\'e}r}(1999)}]{Cramer1999}%
  \BibitemOpen
  \bibfield  {author} {\bibinfo {author} {\bibfnamefont {Harald}\ \bibnamefont
  {Cram{\'e}r}},\ }\href@noop {} {\emph {\bibinfo {title} {Mathematical methods
  of statistics}}},\ Vol.~\bibinfo {volume} {9}\ (\bibinfo  {publisher}
  {Princeton university press},\ \bibinfo {year} {1999})\BibitemShut {NoStop}%
\bibitem [{\citenamefont {Rohatgi}\ and\ \citenamefont
  {Saleh}(2000)}]{Rohatgi2000}%
  \BibitemOpen
  \bibfield  {author} {\bibinfo {author} {\bibfnamefont {Vijay~K}\ \bibnamefont
  {Rohatgi}}\ and\ \bibinfo {author} {\bibfnamefont {AK~Md~Ehsanes}\
  \bibnamefont {Saleh}},\ }\href {\doibase 10.1002/9781118165676} {\emph
  {\bibinfo {title} {An introduction to probability and statistics}}}\
  (\bibinfo  {publisher} {John Wiley \& Sons},\ \bibinfo {year}
  {2000})\BibitemShut {NoStop}%
\bibitem [{\citenamefont {Helstrom}(1969)}]{Helstrom1969}%
  \BibitemOpen
  \bibfield  {author} {\bibinfo {author} {\bibfnamefont {Carl~W.}\ \bibnamefont
  {Helstrom}},\ }\bibfield  {title} {\enquote {\bibinfo {title} {Quantum
  detection and estimation theory},}\ }\href {\doibase 10.1007/BF01007479}
  {\bibfield  {journal} {\bibinfo  {journal} {Journal of Statistical Physics}\
  }\textbf {\bibinfo {volume} {1}},\ \bibinfo {pages} {231--252} (\bibinfo
  {year} {1969})}\BibitemShut {NoStop}%
\bibitem [{\citenamefont {Holevo}(2011)}]{Holevo2011}%
  \BibitemOpen
  \bibfield  {author} {\bibinfo {author} {\bibfnamefont {A.S.}\ \bibnamefont
  {Holevo}},\ }\href {https://books.google.co.uk/books?id=l7AIDhbWrTIC} {\emph
  {\bibinfo {title} {Probabilistic and Statistical Aspects of Quantum
  Theory}}},\ Publications of the Scuola Normale Superiore\ (\bibinfo
  {publisher} {Scuola Normale Superiore},\ \bibinfo {year} {2011})\BibitemShut
  {NoStop}%
\bibitem [{\citenamefont {Scott}\ \emph {et~al.}(2020)\citenamefont {Scott},
  \citenamefont {Branford}, \citenamefont {Westerberg}, \citenamefont {Leach},\
  and\ \citenamefont {Gauger}}]{Scott2020}%
  \BibitemOpen
  \bibfield  {author} {\bibinfo {author} {\bibfnamefont {Hamish}\ \bibnamefont
  {Scott}}, \bibinfo {author} {\bibfnamefont {Dominic}\ \bibnamefont
  {Branford}}, \bibinfo {author} {\bibfnamefont {Niclas}\ \bibnamefont
  {Westerberg}}, \bibinfo {author} {\bibfnamefont {Jonathan}\ \bibnamefont
  {Leach}}, \ and\ \bibinfo {author} {\bibfnamefont {Erik~M.}\ \bibnamefont
  {Gauger}},\ }\bibfield  {title} {\enquote {\bibinfo {title} {Beyond
  coincidence in hong-ou-mandel interferometry},}\ }\href {\doibase
  10.1103/PhysRevA.102.033714} {\bibfield  {journal} {\bibinfo  {journal}
  {Phys. Rev. A}\ }\textbf {\bibinfo {volume} {102}},\ \bibinfo {pages}
  {033714} (\bibinfo {year} {2020})}\BibitemShut {NoStop}%
\bibitem [{\citenamefont {Fabre}\ and\ \citenamefont
  {Felicetti}(2021)}]{Fabre2021}%
  \BibitemOpen
  \bibfield  {author} {\bibinfo {author} {\bibfnamefont {N.}~\bibnamefont
  {Fabre}}\ and\ \bibinfo {author} {\bibfnamefont {S.}~\bibnamefont
  {Felicetti}},\ }\bibfield  {title} {\enquote {\bibinfo {title} {Parameter
  estimation of time and frequency shifts with generalized hong-ou-mandel
  interferometry},}\ }\href {\doibase 10.1103/PhysRevA.104.022208} {\bibfield
  {journal} {\bibinfo  {journal} {Phys. Rev. A}\ }\textbf {\bibinfo {volume}
  {104}},\ \bibinfo {pages} {022208} (\bibinfo {year} {2021})}\BibitemShut
  {NoStop}%
\bibitem [{\citenamefont {Johnson}\ \emph {et~al.}(2023)\citenamefont
  {Johnson}, \citenamefont {Lualdi}, \citenamefont {Conrad}, \citenamefont
  {Arnold}, \citenamefont {Vayninger},\ and\ \citenamefont
  {Kwiat}}]{Johnson2023}%
  \BibitemOpen
  \bibfield  {author} {\bibinfo {author} {\bibfnamefont {Spencer~J.}\
  \bibnamefont {Johnson}}, \bibinfo {author} {\bibfnamefont {Colin~P.}\
  \bibnamefont {Lualdi}}, \bibinfo {author} {\bibfnamefont {Andrew~P.}\
  \bibnamefont {Conrad}}, \bibinfo {author} {\bibfnamefont {Nathan~T.}\
  \bibnamefont {Arnold}}, \bibinfo {author} {\bibfnamefont {Michael}\
  \bibnamefont {Vayninger}}, \ and\ \bibinfo {author} {\bibfnamefont {Paul~G.}\
  \bibnamefont {Kwiat}},\ }\bibfield  {title} {\enquote {\bibinfo {title}
  {{Toward vibration measurement via frequency-entangled two-photon
  interferometry}},}\ }in\ \href {\doibase 10.1117/12.2650820} {\emph {\bibinfo
  {booktitle} {Quantum Sensing, Imaging, and Precision Metrology}}},\ Vol.\
  \bibinfo {volume} {12447}\ (\bibinfo  {publisher} {SPIE},\ \bibinfo {year}
  {2023})\ p.\ \bibinfo {pages} {124471C}\BibitemShut {NoStop}%
\bibitem [{\citenamefont {Legero}\ \emph {et~al.}(2003)\citenamefont {Legero},
  \citenamefont {Wilk}, \citenamefont {Kuhn},\ and\ \citenamefont
  {Rempe}}]{Legero2003}%
  \BibitemOpen
  \bibfield  {author} {\bibinfo {author} {\bibfnamefont {T.}~\bibnamefont
  {Legero}}, \bibinfo {author} {\bibfnamefont {T.}~\bibnamefont {Wilk}},
  \bibinfo {author} {\bibfnamefont {A.}~\bibnamefont {Kuhn}}, \ and\ \bibinfo
  {author} {\bibfnamefont {G.}~\bibnamefont {Rempe}},\ }\bibfield  {title}
  {\enquote {\bibinfo {title} {Time-resolved two-photon quantum
  interference},}\ }\href {\doibase 10.1007/s00340-003-1337-x} {\bibfield
  {journal} {\bibinfo  {journal} {Applied Physics B}\ }\textbf {\bibinfo
  {volume} {77}},\ \bibinfo {pages} {797--802} (\bibinfo {year}
  {2003})}\BibitemShut {NoStop}%
\bibitem [{\citenamefont {Legero}\ \emph {et~al.}(2004)\citenamefont {Legero},
  \citenamefont {Wilk}, \citenamefont {Hennrich}, \citenamefont {Rempe},\ and\
  \citenamefont {Kuhn}}]{Legero2004}%
  \BibitemOpen
  \bibfield  {author} {\bibinfo {author} {\bibfnamefont {Thomas}\ \bibnamefont
  {Legero}}, \bibinfo {author} {\bibfnamefont {Tatjana}\ \bibnamefont {Wilk}},
  \bibinfo {author} {\bibfnamefont {Markus}\ \bibnamefont {Hennrich}}, \bibinfo
  {author} {\bibfnamefont {Gerhard}\ \bibnamefont {Rempe}}, \ and\ \bibinfo
  {author} {\bibfnamefont {Axel}\ \bibnamefont {Kuhn}},\ }\bibfield  {title}
  {\enquote {\bibinfo {title} {Quantum beat of two single photons},}\ }\href
  {\doibase 10.1103/PhysRevLett.93.070503} {\bibfield  {journal} {\bibinfo
  {journal} {Phys. Rev. Lett.}\ }\textbf {\bibinfo {volume} {93}},\ \bibinfo
  {pages} {070503} (\bibinfo {year} {2004})}\BibitemShut {NoStop}%
\bibitem [{\citenamefont {Tamma}\ and\ \citenamefont
  {Laibacher}(2015)}]{Tamma2015}%
  \BibitemOpen
  \bibfield  {author} {\bibinfo {author} {\bibfnamefont {Vincenzo}\
  \bibnamefont {Tamma}}\ and\ \bibinfo {author} {\bibfnamefont {Simon}\
  \bibnamefont {Laibacher}},\ }\bibfield  {title} {\enquote {\bibinfo {title}
  {Multiboson correlation interferometry with arbitrary single-photon pure
  states},}\ }\href {\doibase 10.1103/PhysRevLett.114.243601} {\bibfield
  {journal} {\bibinfo  {journal} {Phys. Rev. Lett.}\ }\textbf {\bibinfo
  {volume} {114}},\ \bibinfo {pages} {243601} (\bibinfo {year}
  {2015})}\BibitemShut {NoStop}%
\bibitem [{\citenamefont {Jin}\ \emph {et~al.}(2015)\citenamefont {Jin},
  \citenamefont {Gerrits}, \citenamefont {Fujiwara}, \citenamefont
  {Wakabayashi}, \citenamefont {Yamashita}, \citenamefont {Miki}, \citenamefont
  {Terai}, \citenamefont {Shimizu}, \citenamefont {Takeoka},\ and\
  \citenamefont {Sasaki}}]{Jin2015}%
  \BibitemOpen
  \bibfield  {author} {\bibinfo {author} {\bibfnamefont {Rui-Bo}\ \bibnamefont
  {Jin}}, \bibinfo {author} {\bibfnamefont {Thomas}\ \bibnamefont {Gerrits}},
  \bibinfo {author} {\bibfnamefont {Mikio}\ \bibnamefont {Fujiwara}}, \bibinfo
  {author} {\bibfnamefont {Ryota}\ \bibnamefont {Wakabayashi}}, \bibinfo
  {author} {\bibfnamefont {Taro}\ \bibnamefont {Yamashita}}, \bibinfo {author}
  {\bibfnamefont {Shigehito}\ \bibnamefont {Miki}}, \bibinfo {author}
  {\bibfnamefont {Hirotaka}\ \bibnamefont {Terai}}, \bibinfo {author}
  {\bibfnamefont {Ryosuke}\ \bibnamefont {Shimizu}}, \bibinfo {author}
  {\bibfnamefont {Masahiro}\ \bibnamefont {Takeoka}}, \ and\ \bibinfo {author}
  {\bibfnamefont {Masahide}\ \bibnamefont {Sasaki}},\ }\bibfield  {title}
  {\enquote {\bibinfo {title} {Spectrally resolved hong-ou-mandel interference
  between independent photon sources},}\ }\href {\doibase 10.1364/OE.23.028836}
  {\bibfield  {journal} {\bibinfo  {journal} {Opt. Express}\ }\textbf {\bibinfo
  {volume} {23}},\ \bibinfo {pages} {28836--28848} (\bibinfo {year}
  {2015})}\BibitemShut {NoStop}%
\bibitem [{\citenamefont {Yepiz-Graciano}\ \emph {et~al.}(2020)\citenamefont
  {Yepiz-Graciano}, \citenamefont {Mart\'{i}nez}, \citenamefont {Lopez-Mago},
  \citenamefont {Cruz-Ramirez},\ and\ \citenamefont
  {U'Ren}}]{Yepiz-Graciano2020}%
  \BibitemOpen
  \bibfield  {author} {\bibinfo {author} {\bibfnamefont {Pablo}\ \bibnamefont
  {Yepiz-Graciano}}, \bibinfo {author} {\bibfnamefont {Al\'{i} Michel~Angulo}\
  \bibnamefont {Mart\'{i}nez}}, \bibinfo {author} {\bibfnamefont {Dorilian}\
  \bibnamefont {Lopez-Mago}}, \bibinfo {author} {\bibfnamefont {Hector}\
  \bibnamefont {Cruz-Ramirez}}, \ and\ \bibinfo {author} {\bibfnamefont
  {Alfred~B.}\ \bibnamefont {U'Ren}},\ }\bibfield  {title} {\enquote {\bibinfo
  {title} {Spectrally resolved hong-ou-mandel interferometry for
  quantum-optical coherence tomography},}\ }\href {\doibase 10.1364/PRJ.388693}
  {\bibfield  {journal} {\bibinfo  {journal} {Photon. Res.}\ }\textbf {\bibinfo
  {volume} {8}},\ \bibinfo {pages} {1023--1034} (\bibinfo {year}
  {2020})}\BibitemShut {NoStop}%
\bibitem [{\citenamefont {Triggiani}\ \emph {et~al.}(2023)\citenamefont
  {Triggiani}, \citenamefont {Psaroudis},\ and\ \citenamefont
  {Tamma}}]{Triggiani2023}%
  \BibitemOpen
  \bibfield  {author} {\bibinfo {author} {\bibfnamefont {Danilo}\ \bibnamefont
  {Triggiani}}, \bibinfo {author} {\bibfnamefont {Giorgos}\ \bibnamefont
  {Psaroudis}}, \ and\ \bibinfo {author} {\bibfnamefont {Vincenzo}\
  \bibnamefont {Tamma}},\ }\bibfield  {title} {\enquote {\bibinfo {title}
  {Ultimate quantum sensitivity in the estimation of the delay between two
  interfering photons through frequency-resolving sampling},}\ }\href {\doibase
  10.1103/PhysRevApplied.19.044068} {\bibfield  {journal} {\bibinfo  {journal}
  {Phys. Rev. Appl.}\ }\textbf {\bibinfo {volume} {19}},\ \bibinfo {pages}
  {044068} (\bibinfo {year} {2023})}\BibitemShut {NoStop}%
\bibitem [{\citenamefont {Wang}\ \emph {et~al.}(2018)\citenamefont {Wang},
  \citenamefont {Jing}, \citenamefont {Sun}, \citenamefont {Yang},
  \citenamefont {Yu}, \citenamefont {Tamma}, \citenamefont {Bao},\ and\
  \citenamefont {Pan}}]{Wang2018}%
  \BibitemOpen
  \bibfield  {author} {\bibinfo {author} {\bibfnamefont {Xu-Jie}\ \bibnamefont
  {Wang}}, \bibinfo {author} {\bibfnamefont {Bo}~\bibnamefont {Jing}}, \bibinfo
  {author} {\bibfnamefont {Peng-Fei}\ \bibnamefont {Sun}}, \bibinfo {author}
  {\bibfnamefont {Chao-Wei}\ \bibnamefont {Yang}}, \bibinfo {author}
  {\bibfnamefont {Yong}\ \bibnamefont {Yu}}, \bibinfo {author} {\bibfnamefont
  {Vincenzo}\ \bibnamefont {Tamma}}, \bibinfo {author} {\bibfnamefont
  {Xiao-Hui}\ \bibnamefont {Bao}}, \ and\ \bibinfo {author} {\bibfnamefont
  {Jian-Wei}\ \bibnamefont {Pan}},\ }\bibfield  {title} {\enquote {\bibinfo
  {title} {Experimental time-resolved interference with multiple photons of
  different colors},}\ }\href {\doibase 10.1103/PhysRevLett.121.080501}
  {\bibfield  {journal} {\bibinfo  {journal} {Phys. Rev. Lett.}\ }\textbf
  {\bibinfo {volume} {121}},\ \bibinfo {pages} {080501} (\bibinfo {year}
  {2018})}\BibitemShut {NoStop}%
\bibitem [{\citenamefont {Hiemstra}\ \emph {et~al.}(2020)\citenamefont
  {Hiemstra}, \citenamefont {Parker}, \citenamefont {Humphreys}, \citenamefont
  {Tiedau}, \citenamefont {Beck}, \citenamefont {Karpi\ifmmode~\acute{n}\else
  \'{n}\fi{}ski}, \citenamefont {Smith}, \citenamefont {Eckstein},
  \citenamefont {Kolthammer},\ and\ \citenamefont {Walmsley}}]{Hiemstra2020}%
  \BibitemOpen
  \bibfield  {author} {\bibinfo {author} {\bibfnamefont {T.}~\bibnamefont
  {Hiemstra}}, \bibinfo {author} {\bibfnamefont {T.F.}\ \bibnamefont {Parker}},
  \bibinfo {author} {\bibfnamefont {P.}~\bibnamefont {Humphreys}}, \bibinfo
  {author} {\bibfnamefont {J.}~\bibnamefont {Tiedau}}, \bibinfo {author}
  {\bibfnamefont {M.}~\bibnamefont {Beck}}, \bibinfo {author} {\bibfnamefont
  {M.}~\bibnamefont {Karpi\ifmmode~\acute{n}\else \'{n}\fi{}ski}}, \bibinfo
  {author} {\bibfnamefont {B.J.}\ \bibnamefont {Smith}}, \bibinfo {author}
  {\bibfnamefont {A.}~\bibnamefont {Eckstein}}, \bibinfo {author}
  {\bibfnamefont {W.S.}\ \bibnamefont {Kolthammer}}, \ and\ \bibinfo {author}
  {\bibfnamefont {I.A.}\ \bibnamefont {Walmsley}},\ }\bibfield  {title}
  {\enquote {\bibinfo {title} {Pure single photons from scalable frequency
  multiplexing},}\ }\href {\doibase 10.1103/PhysRevApplied.14.014052}
  {\bibfield  {journal} {\bibinfo  {journal} {Phys. Rev. Applied}\ }\textbf
  {\bibinfo {volume} {14}},\ \bibinfo {pages} {014052} (\bibinfo {year}
  {2020})}\BibitemShut {NoStop}%
\bibitem [{\citenamefont {Prakash}\ \emph {et~al.}(2021)\citenamefont
  {Prakash}, \citenamefont {Sierant},\ and\ \citenamefont
  {Mitchell}}]{Prakash2021}%
  \BibitemOpen
  \bibfield  {author} {\bibinfo {author} {\bibfnamefont {Vindhiya}\
  \bibnamefont {Prakash}}, \bibinfo {author} {\bibfnamefont {Aleksandra}\
  \bibnamefont {Sierant}}, \ and\ \bibinfo {author} {\bibfnamefont {Morgan~W.}\
  \bibnamefont {Mitchell}},\ }\bibfield  {title} {\enquote {\bibinfo {title}
  {Autoheterodyne characterization of narrow-band photon pairs},}\ }\href
  {\doibase 10.1103/PhysRevLett.127.043601} {\bibfield  {journal} {\bibinfo
  {journal} {Phys. Rev. Lett.}\ }\textbf {\bibinfo {volume} {127}},\ \bibinfo
  {pages} {043601} (\bibinfo {year} {2021})}\BibitemShut {NoStop}%
\bibitem [{\citenamefont {Ou}\ and\ \citenamefont {Mandel}(1989)}]{Ou1988}%
  \BibitemOpen
  \bibfield  {author} {\bibinfo {author} {\bibfnamefont {Z.~Y.}\ \bibnamefont
  {Ou}}\ and\ \bibinfo {author} {\bibfnamefont {L.}~\bibnamefont {Mandel}},\
  }\bibfield  {title} {\enquote {\bibinfo {title} {Further evidence of
  nonclassical behavior in optical interference},}\ }\href {\doibase
  10.1103/PhysRevLett.62.2941} {\bibfield  {journal} {\bibinfo  {journal}
  {Phys. Rev. Lett.}\ }\textbf {\bibinfo {volume} {62}},\ \bibinfo {pages}
  {2941--2944} (\bibinfo {year} {1989})}\BibitemShut {NoStop}%
\bibitem [{\citenamefont {Kim}\ \emph {et~al.}(2006)\citenamefont {Kim},
  \citenamefont {Kwon}, \citenamefont {Kim},\ and\ \citenamefont
  {Kim}}]{Kim2006}%
  \BibitemOpen
  \bibfield  {author} {\bibinfo {author} {\bibfnamefont {Heonoh}\ \bibnamefont
  {Kim}}, \bibinfo {author} {\bibfnamefont {Osung}\ \bibnamefont {Kwon}},
  \bibinfo {author} {\bibfnamefont {Wonsik}\ \bibnamefont {Kim}}, \ and\
  \bibinfo {author} {\bibfnamefont {Taesoo}\ \bibnamefont {Kim}},\ }\bibfield
  {title} {\enquote {\bibinfo {title} {Spatial two-photon interference in a
  hong-ou-mandel interferometer},}\ }\href {\doibase
  10.1103/PhysRevA.73.023820} {\bibfield  {journal} {\bibinfo  {journal} {Phys.
  Rev. A}\ }\textbf {\bibinfo {volume} {73}},\ \bibinfo {pages} {023820}
  (\bibinfo {year} {2006})}\BibitemShut {NoStop}%
\bibitem [{\citenamefont {Lee}\ and\ \citenamefont {van
  Exter}(2006)}]{Lee2006}%
  \BibitemOpen
  \bibfield  {author} {\bibinfo {author} {\bibfnamefont {P.~S.~K.}\
  \bibnamefont {Lee}}\ and\ \bibinfo {author} {\bibfnamefont {M.~P.}\
  \bibnamefont {van Exter}},\ }\bibfield  {title} {\enquote {\bibinfo {title}
  {Spatial labeling in a two-photon interferometer},}\ }\href {\doibase
  10.1103/PhysRevA.73.063827} {\bibfield  {journal} {\bibinfo  {journal} {Phys.
  Rev. A}\ }\textbf {\bibinfo {volume} {73}},\ \bibinfo {pages} {063827}
  (\bibinfo {year} {2006})}\BibitemShut {NoStop}%
\bibitem [{\citenamefont {Devaux}\ \emph {et~al.}(2020)\citenamefont {Devaux},
  \citenamefont {Mosset}, \citenamefont {Moreau},\ and\ \citenamefont
  {Lantz}}]{Devaux2020}%
  \BibitemOpen
  \bibfield  {author} {\bibinfo {author} {\bibfnamefont {Fabrice}\ \bibnamefont
  {Devaux}}, \bibinfo {author} {\bibfnamefont {Alexis}\ \bibnamefont {Mosset}},
  \bibinfo {author} {\bibfnamefont {Paul-Antoine}\ \bibnamefont {Moreau}}, \
  and\ \bibinfo {author} {\bibfnamefont {Eric}\ \bibnamefont {Lantz}},\
  }\bibfield  {title} {\enquote {\bibinfo {title} {Imaging spatiotemporal
  hong-ou-mandel interference of biphoton states of extremely high schmidt
  number},}\ }\href {\doibase 10.1103/PhysRevX.10.031031} {\bibfield  {journal}
  {\bibinfo  {journal} {Phys. Rev. X}\ }\textbf {\bibinfo {volume} {10}},\
  \bibinfo {pages} {031031} (\bibinfo {year} {2020})}\BibitemShut {NoStop}%
\bibitem [{\citenamefont {Hanne}\ \emph {et~al.}(2015)\citenamefont {Hanne},
  \citenamefont {Falk}, \citenamefont {G{\"o}rlitz}, \citenamefont {Hoyer},
  \citenamefont {Engelhardt}, \citenamefont {Sahl},\ and\ \citenamefont
  {Hell}}]{Hanne2015}%
  \BibitemOpen
  \bibfield  {author} {\bibinfo {author} {\bibfnamefont {Janina}\ \bibnamefont
  {Hanne}}, \bibinfo {author} {\bibfnamefont {Henning~J.}\ \bibnamefont
  {Falk}}, \bibinfo {author} {\bibfnamefont {Frederik}\ \bibnamefont
  {G{\"o}rlitz}}, \bibinfo {author} {\bibfnamefont {Patrick}\ \bibnamefont
  {Hoyer}}, \bibinfo {author} {\bibfnamefont {Johann}\ \bibnamefont
  {Engelhardt}}, \bibinfo {author} {\bibfnamefont {Steffen~J.}\ \bibnamefont
  {Sahl}}, \ and\ \bibinfo {author} {\bibfnamefont {Stefan~W.}\ \bibnamefont
  {Hell}},\ }\bibfield  {title} {\enquote {\bibinfo {title} {Sted nanoscopy
  with fluorescent quantum dots},}\ }\href {\doibase 10.1038/ncomms8127}
  {\bibfield  {journal} {\bibinfo  {journal} {Nature Communications}\ }\textbf
  {\bibinfo {volume} {6}},\ \bibinfo {pages} {7127} (\bibinfo {year}
  {2015})}\BibitemShut {NoStop}%
\bibitem [{\citenamefont {Bruschini}\ \emph {et~al.}(2019)\citenamefont
  {Bruschini}, \citenamefont {Homulle}, \citenamefont {Antolovic},
  \citenamefont {Burri},\ and\ \citenamefont {Charbon}}]{Bruschini2019}%
  \BibitemOpen
  \bibfield  {author} {\bibinfo {author} {\bibfnamefont {Claudio}\ \bibnamefont
  {Bruschini}}, \bibinfo {author} {\bibfnamefont {Harald}\ \bibnamefont
  {Homulle}}, \bibinfo {author} {\bibfnamefont {Ivan~Michel}\ \bibnamefont
  {Antolovic}}, \bibinfo {author} {\bibfnamefont {Samuel}\ \bibnamefont
  {Burri}}, \ and\ \bibinfo {author} {\bibfnamefont {Edoardo}\ \bibnamefont
  {Charbon}},\ }\bibfield  {title} {\enquote {\bibinfo {title} {Single-photon
  avalanche diode imagers in biophotonics: review and outlook},}\ }\href
  {\doibase 10.1038/s41377-019-0191-5} {\bibfield  {journal} {\bibinfo
  {journal} {Light: Science {\&} Applications}\ }\textbf {\bibinfo {volume}
  {8}},\ \bibinfo {pages} {87} (\bibinfo {year} {2019})}\BibitemShut {NoStop}%
\bibitem [{\citenamefont {Urban}\ \emph {et~al.}(2021)\citenamefont {Urban},
  \citenamefont {Chiang}, \citenamefont {Hammond}, \citenamefont {Cogan},
  \citenamefont {Litzburg}, \citenamefont {Burke}, \citenamefont {Stern},
  \citenamefont {Gelbard}, \citenamefont {Nilsson},\ and\ \citenamefont
  {Krauss}}]{Urban2021}%
  \BibitemOpen
  \bibfield  {author} {\bibinfo {author} {\bibfnamefont {Jennifer~M.}\
  \bibnamefont {Urban}}, \bibinfo {author} {\bibfnamefont {Wesley}\
  \bibnamefont {Chiang}}, \bibinfo {author} {\bibfnamefont {Jennetta~W.}\
  \bibnamefont {Hammond}}, \bibinfo {author} {\bibfnamefont {Nicole M.~B.}\
  \bibnamefont {Cogan}}, \bibinfo {author} {\bibfnamefont {Angela}\
  \bibnamefont {Litzburg}}, \bibinfo {author} {\bibfnamefont {Rebeckah}\
  \bibnamefont {Burke}}, \bibinfo {author} {\bibfnamefont {Harry~A.}\
  \bibnamefont {Stern}}, \bibinfo {author} {\bibfnamefont {Harris~A.}\
  \bibnamefont {Gelbard}}, \bibinfo {author} {\bibfnamefont {Bradley~L.}\
  \bibnamefont {Nilsson}}, \ and\ \bibinfo {author} {\bibfnamefont {Todd~D.}\
  \bibnamefont {Krauss}},\ }\bibfield  {title} {\enquote {\bibinfo {title}
  {Quantum dots for improved single-molecule localization microscopy},}\ }\href
  {\doibase 10.1021/acs.jpcb.0c11545} {\bibfield  {journal} {\bibinfo
  {journal} {The Journal of Physical Chemistry B}\ }\textbf {\bibinfo {volume}
  {125}},\ \bibinfo {pages} {2566--2576} (\bibinfo {year} {2021})},\ \bibinfo
  {note} {pMID: 33683893},\ \Eprint
  {http://arxiv.org/abs/https://doi.org/10.1021/acs.jpcb.0c11545}
  {https://doi.org/10.1021/acs.jpcb.0c11545} \BibitemShut {NoStop}%
\bibitem [{\citenamefont {Laibacher}\ and\ \citenamefont
  {Tamma}(2015)}]{Laibacher2015}%
  \BibitemOpen
  \bibfield  {author} {\bibinfo {author} {\bibfnamefont {Simon}\ \bibnamefont
  {Laibacher}}\ and\ \bibinfo {author} {\bibfnamefont {Vincenzo}\ \bibnamefont
  {Tamma}},\ }\bibfield  {title} {\enquote {\bibinfo {title} {From the physics
  to the computational complexity of multiboson correlation interference},}\
  }\href {\doibase 10.1103/PhysRevLett.115.243605} {\bibfield  {journal}
  {\bibinfo  {journal} {Phys. Rev. Lett.}\ }\textbf {\bibinfo {volume} {115}},\
  \bibinfo {pages} {243605} (\bibinfo {year} {2015})}\BibitemShut {NoStop}%
\bibitem [{\citenamefont {Tamma}\ and\ \citenamefont
  {Laibacher}(2016)}]{Tamma2016}%
  \BibitemOpen
  \bibfield  {author} {\bibinfo {author} {\bibfnamefont {Vincenzo}\
  \bibnamefont {Tamma}}\ and\ \bibinfo {author} {\bibfnamefont {Simon}\
  \bibnamefont {Laibacher}},\ }\bibfield  {title} {\enquote {\bibinfo {title}
  {Multi-boson correlation sampling},}\ }\href {\doibase
  10.1007/s11128-015-1177-8} {\bibfield  {journal} {\bibinfo  {journal}
  {Quantum Information Processing}\ }\textbf {\bibinfo {volume} {15}},\
  \bibinfo {pages} {1241--1262} (\bibinfo {year} {2016})}\BibitemShut {NoStop}%
\bibitem [{\citenamefont {Tamma}\ and\ \citenamefont
  {Laibacher}(2023)}]{Tamma2023}%
  \BibitemOpen
  \bibfield  {author} {\bibinfo {author} {\bibfnamefont {Vincenzo}\
  \bibnamefont {Tamma}}\ and\ \bibinfo {author} {\bibfnamefont {Simon}\
  \bibnamefont {Laibacher}},\ }\bibfield  {title} {\enquote {\bibinfo {title}
  {Scattershot multiboson correlation sampling with random photonic inner-mode
  multiplexing},}\ }\href {\doibase 10.1140/epjp/s13360-023-03941-2} {\bibfield
   {journal} {\bibinfo  {journal} {The European Physical Journal Plus}\
  }\textbf {\bibinfo {volume} {138}},\ \bibinfo {pages} {335} (\bibinfo {year}
  {2023})}\BibitemShut {NoStop}%
\bibitem [{\citenamefont {Ober}\ \emph {et~al.}(2004)\citenamefont {Ober},
  \citenamefont {Ram},\ and\ \citenamefont {Ward}}]{Ober2004}%
  \BibitemOpen
  \bibfield  {author} {\bibinfo {author} {\bibfnamefont {Raimund~J.}\
  \bibnamefont {Ober}}, \bibinfo {author} {\bibfnamefont {Sripad}\ \bibnamefont
  {Ram}}, \ and\ \bibinfo {author} {\bibfnamefont {E.~Sally}\ \bibnamefont
  {Ward}},\ }\bibfield  {title} {\enquote {\bibinfo {title} {Localization
  accuracy in single-molecule microscopy},}\ }\href {\doibase
  https://doi.org/10.1016/S0006-3495(04)74193-4} {\bibfield  {journal}
  {\bibinfo  {journal} {Biophysical Journal}\ }\textbf {\bibinfo {volume}
  {86}},\ \bibinfo {pages} {1185--1200} (\bibinfo {year} {2004})}\BibitemShut
  {NoStop}%
\bibitem [{\citenamefont {Lelek}\ \emph {et~al.}(2021)\citenamefont {Lelek},
  \citenamefont {Gyparaki}, \citenamefont {Beliu}, \citenamefont {Schueder},
  \citenamefont {Griffi{\'e}}, \citenamefont {Manley}, \citenamefont
  {Jungmann}, \citenamefont {Sauer}, \citenamefont {Lakadamyali},\ and\
  \citenamefont {Zimmer}}]{Lelek2021}%
  \BibitemOpen
  \bibfield  {author} {\bibinfo {author} {\bibfnamefont {Micka{\"e}l}\
  \bibnamefont {Lelek}}, \bibinfo {author} {\bibfnamefont {Melina~T.}\
  \bibnamefont {Gyparaki}}, \bibinfo {author} {\bibfnamefont {Gerti}\
  \bibnamefont {Beliu}}, \bibinfo {author} {\bibfnamefont {Florian}\
  \bibnamefont {Schueder}}, \bibinfo {author} {\bibfnamefont {Juliette}\
  \bibnamefont {Griffi{\'e}}}, \bibinfo {author} {\bibfnamefont {Suliana}\
  \bibnamefont {Manley}}, \bibinfo {author} {\bibfnamefont {Ralf}\ \bibnamefont
  {Jungmann}}, \bibinfo {author} {\bibfnamefont {Markus}\ \bibnamefont
  {Sauer}}, \bibinfo {author} {\bibfnamefont {Melike}\ \bibnamefont
  {Lakadamyali}}, \ and\ \bibinfo {author} {\bibfnamefont {Christophe}\
  \bibnamefont {Zimmer}},\ }\bibfield  {title} {\enquote {\bibinfo {title}
  {Single-molecule localization microscopy},}\ }\href {\doibase
  10.1038/s43586-021-00038-x} {\bibfield  {journal} {\bibinfo  {journal}
  {Nature Reviews Methods Primers}\ }\textbf {\bibinfo {volume} {1}},\ \bibinfo
  {pages} {39} (\bibinfo {year} {2021})}\BibitemShut {NoStop}%
\bibitem [{\citenamefont {Salazar-Serrano}\ \emph {et~al.}(2015)\citenamefont
  {Salazar-Serrano}, \citenamefont {Valencia},\ and\ \citenamefont
  {Torres}}]{Serrano2015}%
  \BibitemOpen
  \bibfield  {author} {\bibinfo {author} {\bibfnamefont {Luis~José}\
  \bibnamefont {Salazar-Serrano}}, \bibinfo {author} {\bibfnamefont
  {Alejandra}\ \bibnamefont {Valencia}}, \ and\ \bibinfo {author}
  {\bibfnamefont {Juan~P.}\ \bibnamefont {Torres}},\ }\bibfield  {title}
  {\enquote {\bibinfo {title} {{Tunable beam displacer}},}\ }\href {\doibase
  10.1063/1.4914834} {\bibfield  {journal} {\bibinfo  {journal} {Review of
  Scientific Instruments}\ }\textbf {\bibinfo {volume} {86}} (\bibinfo {year}
  {2015}),\ 10.1063/1.4914834},\ \bibinfo {note} {033109},\ \Eprint
  {http://arxiv.org/abs/https://pubs.aip.org/aip/rsi/article-pdf/doi/10.1063/1.4914834/15818465/033109\_1\_online.pdf}
  {https://pubs.aip.org/aip/rsi/article-pdf/doi/10.1063/1.4914834/15818465/033109\_1\_online.pdf}
  \BibitemShut {NoStop}%
\bibitem [{Sup()}]{Suppl}%
  \BibitemOpen
  \href@noop {} {}\bibinfo {note} {See Appendices (which include
  \cite{Glauber1963}) for the derivations of
  \cref{eq:Prob,eq:QFI,eq:FINu,eq:FIContr}, the maximum likelihood function,
  and the FI for $\sigma_k\Delta x\gg 1$, for non-resolved measurements, and
  for the observation of only single-camera or two-camera events.}\BibitemShut
  {Stop}%
\bibitem [{\citenamefont {Glauber}(1963)}]{Glauber1963}%
  \BibitemOpen
  \bibfield  {author} {\bibinfo {author} {\bibfnamefont {Roy~J.}\ \bibnamefont
  {Glauber}},\ }\bibfield  {title} {\enquote {\bibinfo {title} {The quantum
  theory of optical coherence},}\ }\href {\doibase 10.1103/PhysRev.130.2529}
  {\bibfield  {journal} {\bibinfo  {journal} {Phys. Rev.}\ }\textbf {\bibinfo
  {volume} {130}},\ \bibinfo {pages} {2529--2539} (\bibinfo {year}
  {1963})}\BibitemShut {NoStop}%
\bibitem [{\citenamefont {Paris}(2009)}]{Paris2009}%
  \BibitemOpen
  \bibfield  {author} {\bibinfo {author} {\bibfnamefont {Matteo G.~A.}\
  \bibnamefont {Paris}},\ }\bibfield  {title} {\enquote {\bibinfo {title}
  {Quantum estimation for quantum technology},}\ }\href {\doibase
  10.1142/S0219749909004839} {\bibfield  {journal} {\bibinfo  {journal}
  {International Journal of Quantum Information}\ }\textbf {\bibinfo {volume}
  {07}},\ \bibinfo {pages} {125--137} (\bibinfo {year} {2009})},\ \Eprint
  {http://arxiv.org/abs/https://doi.org/10.1142/S0219749909004839}
  {https://doi.org/10.1142/S0219749909004839} \BibitemShut {NoStop}%
\bibitem [{\citenamefont {Liu}\ \emph {et~al.}(2019)\citenamefont {Liu},
  \citenamefont {Yuan}, \citenamefont {Lu},\ and\ \citenamefont
  {Wang}}]{Liu2020}%
  \BibitemOpen
  \bibfield  {author} {\bibinfo {author} {\bibfnamefont {Jing}\ \bibnamefont
  {Liu}}, \bibinfo {author} {\bibfnamefont {Haidong}\ \bibnamefont {Yuan}},
  \bibinfo {author} {\bibfnamefont {Xiao-Ming}\ \bibnamefont {Lu}}, \ and\
  \bibinfo {author} {\bibfnamefont {Xiaoguang}\ \bibnamefont {Wang}},\
  }\bibfield  {title} {\enquote {\bibinfo {title} {Quantum fisher information
  matrix and multiparameter estimation},}\ }\href {\doibase
  10.1088/1751-8121/ab5d4d} {\bibfield  {journal} {\bibinfo  {journal} {Journal
  of Physics A: Mathematical and Theoretical}\ }\textbf {\bibinfo {volume}
  {53}},\ \bibinfo {pages} {023001} (\bibinfo {year} {2019})}\BibitemShut
  {NoStop}%
\bibitem [{\citenamefont {Khater}\ \emph {et~al.}(2020)\citenamefont {Khater},
  \citenamefont {Nabi},\ and\ \citenamefont {Hamarneh}}]{Khater2020}%
  \BibitemOpen
  \bibfield  {author} {\bibinfo {author} {\bibfnamefont {Ismail~M.}\
  \bibnamefont {Khater}}, \bibinfo {author} {\bibfnamefont {Ivan~Robert}\
  \bibnamefont {Nabi}}, \ and\ \bibinfo {author} {\bibfnamefont {Ghassan}\
  \bibnamefont {Hamarneh}},\ }\bibfield  {title} {\enquote {\bibinfo {title} {A
  review of super-resolution single-molecule localization microscopy cluster
  analysis and quantification methods},}\ }\href {\doibase
  https://doi.org/10.1016/j.patter.2020.100038} {\bibfield  {journal} {\bibinfo
   {journal} {Patterns}\ }\textbf {\bibinfo {volume} {1}},\ \bibinfo {pages}
  {100038} (\bibinfo {year} {2020})}\BibitemShut {NoStop}%
\bibitem [{\citenamefont {Sahl}\ \emph {et~al.}(2017)\citenamefont {Sahl},
  \citenamefont {Hell},\ and\ \citenamefont {Jakobs}}]{Sahl2017}%
  \BibitemOpen
  \bibfield  {author} {\bibinfo {author} {\bibfnamefont {Steffen~J.}\
  \bibnamefont {Sahl}}, \bibinfo {author} {\bibfnamefont {Stefan~W.}\
  \bibnamefont {Hell}}, \ and\ \bibinfo {author} {\bibfnamefont {Stefan}\
  \bibnamefont {Jakobs}},\ }\bibfield  {title} {\enquote {\bibinfo {title}
  {Fluorescence nanoscopy in cell biology},}\ }\href {\doibase
  10.1038/nrm.2017.71} {\bibfield  {journal} {\bibinfo  {journal} {Nature
  Reviews Molecular Cell Biology}\ }\textbf {\bibinfo {volume} {18}},\ \bibinfo
  {pages} {685--701} (\bibinfo {year} {2017})}\BibitemShut {NoStop}%
\end{thebibliography}%

\appendix
\onecolumngrid
\pagebreak
\section{Joint detection probabilities in Eq.~\eqref{eq:Prob}}
\label{app:Probs}

We will evaluate the probability distribution $P(\Delta k,X)$ in Eq.~\eqref{eq:Prob} associated with the event of detecting the two photons with transverse-momenta differing of $\Delta k=k\PLH-k'$ in the same ($X=B$) and in different ($X=A$) cameras.

Since the detection occurs by means of cameras resolving the position of the photons in the far-field regime, we will first evaluate the second-order correlation functions~\cite{Glauber1963}
\begin{equation}
G^{(2)}_{CC'}(\alpha_1,y\PLH;\alpha_2,y')=\Tr[\ketbra{\Psi}\hat{E}^{-(\alpha_1)}_C(y\PLH)\hat{E}^{-(\alpha_2)}_{C'}(y')\hat{E}^{+(\alpha_1)}_{C}(y\PLH)\hat{E}^{+(\alpha_2)}_{C'}(y')],
\label{eq:G2Def}
\end{equation}
associated with the detection of the two photons, in the state in Eq.~\eqref{eq:InitState} in the main text, in the optical modes (e.g. polarisations) $\alpha_1$ and $\alpha_2$ and at the positions $y\PLH$ and $y'$ of the detectors $C,C'=C_1,C_2$, with $\alpha_i=a,d$, $i=1,2$, where $\hat{a}$ and $\hat{d}$ represent two orthogonal optical modes so that $\hat{b}$ in Eq.~\eqref{eq:InitState} in the main text can be written as $\hat{b}=\sqrt{\nu}\hat{a}+\sqrt{1-\nu}\hat{d}$, and we denote with
\begin{equation}
\hat{E}^{+(\alpha)}_C(y) = \frac{1}{\sqrt{2\pi}}\sum_{S=1,2}\int \d x_S\  g(x_S,S;y,C)\hat{\alpha}_S(x_S),\quad \hat{\alpha}=\hat{a},\hat{d}
\label{eq:ElecFieldD}
\end{equation}
the positive frequency part of the electric field operator at position $y$ of the camera $C$ associated with the optical mode $\hat{\alpha}$, $\hat{E}^-_C(y)=(\hat{E}^+_C(y))^\dag$ the negative frequency part, and
\begin{equation}
g(x,S;y,C) = \frac{\e^{\ii\phi(S,C)}}{\sqrt{2}} \e^{\ii k_0 d}\e^{-\ii \frac{k_0}{d} x y},
\label{eq:Transfer}
\end{equation}
is the Fraunhofer transfer function, describing the propagation of the photon $S=1,2$ from transverse position $x$ through the beam-splitter and up to position $y$ at the detector $C=C_1,C_2$ in the far-field, where $k_0$ is the wavenumber, $d$ is the longitudinal distance travelled by the photons to the detectors, $\phi$ is the phase acquired through the beam-splitter.
In Eq.~\eqref{eq:Transfer}, we have chosen a dimensionless normalisation that guarantees the unitary propagation of the photons through the beam splitter.
By summing the second-order correlation functions in Eq. \eqref{eq:G2Def} for every possible detected optical mode $\alpha_1,\alpha_2$, and for $C=C'$ (same-camera detection) and $C\neq C'$ (two-cameras detection) respectively, we will finally obtain the probability distribution for any value of $\nu$ as shown in Eq.~\eqref{eq:Prob}.

We first introduce the electric field operators
\begin{align}
\label{eq:ElecFields}
\hat{E}_{SC}^{+(\alpha)}(y) &= \frac{1}{\sqrt{2\pi}} \int \d x_S\  g(x_S,S;y,C)\hat{\alpha}_S(x_S),\ S=1,2\ \hat{\alpha}=\hat{a},\hat{d}
\end{align}
associated with the $S$-th photon, so that 
\begin{equation}
\hat{E}^{+(\alpha)}_C(y) = \sum_{S=1,2} \hat{E}^{+(\alpha)}_{SC}(y).
\label{eq:ElecSources}
\end{equation}

We can thus expand the second-order correlation function in Eq.~\eqref{eq:G2Def} by employing Eq.~\eqref{eq:ElecSources}, obtaining
\begin{equation}
G^{(2)}_{CC'}(\alpha_1,y\PLH;\alpha_2,y') = \sum_{\substack{S_1,S_2\\S_3,S_4}=1,2}
\Tr[\ketbra{\Psi}\hat{E}^{-(\alpha_1)}_{S_1C}(y\PLH)\hat{E}^{-(\hat{\alpha}_2)}_{S_2C'}(y')\hat{E}^{+(\hat{\alpha}_1)}_{S_3C}(y\PLH)\hat{E}^{+(\hat{\alpha}_2)}_{S_4C'}(y')].
\label{eq:G2Sum}
\end{equation}
Of the 16 terms in the summation in Eq.~\eqref{eq:G2Sum}, we easily see that only a few survive.
Indeed, since the state $\ket{\Psi}$ has a single photon in each input channel, the terms with $S_1=S_2$ and $S_3= S_4$ vanish.
Moreover, since the positive-frequency part of the electric field annihilates a photon, and the negative-frequency part creates a photon, evaluating the trace in Eq.~\eqref{eq:G2Sum} is equivalent to projecting onto the vacuum $\ketbra{\mathrm{vac}}$, so we have
\begin{align}
G^{(2)}_{CC'}(\alpha_1,y\PLH;\alpha_2,y') &= \sum_{S_1,S_3=1,2}
\mel{\Psi}{\hat{E}^{-(\alpha_1)}_{S_1C}(y\PLH)\hat{E}^{-(\alpha_2)}_{\sigma(S_1)C'}(y')}{\mathrm{vac}}\mel{\mathrm{vac}}{\hat{E}^{+(\alpha_1)}_{S_3C}(y\PLH)\hat{E}^{+(\alpha_2)}_{\sigma(S_3)C'}(y')}{\Psi},\notag\\
&=\abs{\sum_{S_1=1,2}
\mel{\mathrm{vac}}{\hat{E}^{+(\alpha_1)}_{S_1C}(y\PLH)\hat{E}^{+(\alpha_2)}_{\sigma(S_1)C'}(y')}{\Psi}}^2
\label{eq:G2Prod}
\end{align}
where $\sigma(1)=2,\sigma(2)=1$.
We can now proceed with evaluating the scalar products in Eq.~\eqref{eq:G2Prod}.
The generic expression of these scalar products can be expanded into
\begin{multline}
\mel{\mathrm{vac}}{\hat{E}^{+(\alpha)}_{S_1C}(y\PLH)\hat{E}^{+(\beta)}_{\sigma(S_1)C'}(y')}{\Psi}=\frac{1}{2\pi}\int \d x_1\d x_2\d x'_1\d x'_2\  g(x_1,S_1;y\PLH,C)g(x_2,\sigma(S_1);y',C')\\
 \quad\times \psi_1(x'_1)\psi_2(x'_2) \mel{\mathrm{vac}}{\hat{\alpha}_{S_1}(x_1)\hat{\beta}_{\sigma(S_1)}(x_2)\hat{a}^\dag_1(x'_1)\hat{b}^\dag_2(x'_2)}{\mathrm{vac}}.
\label{eq:TermProd}
\end{multline}
We immediately notice that the term for both $\hat{\alpha}=\hat{d}$ and $\hat{\beta}=\hat{d}$ is vanishing since $\comm{\hat{d}_j(y)}{\hat{a}_i^\dag(x)}=0$.
For the same reason, for $\hat{\alpha}=\hat{a}$ and $\hat{\beta}=\hat{d}$, then the term in Eq.~\eqref{eq:TermProd} is non-vanishing only for $S_1=1$, and it reads
\begin{align}
\mel{\mathrm{vac}}{\hat{E}^{+(\hat{a})}_{S_1C}(y\PLH)\hat{E}^{+(\hat{d})}_{\sigma(S_1)C'}(y')}{\Psi}&=\frac{\sqrt{1-\nu}}{2\pi}\int \d x_1\d x_2\d x'_1\d x'_2\  g(x_1,S_1;y\PLH,C)g(x_2,\sigma(S_1);y',C')\\
&\qquad\times\psi_1(x'_1)\psi_2(x'_2)\, \delta_{1S_1}\delta(x_1-x_1')\delta(x_2-x_2')\notag\\
&=\delta_{1S_1}\frac{\sqrt{1-\nu}}{2\pi}\int \d x_1\d x_2\ g(x_1,1;y\PLH,C)g(x_2,2;y',C')\psi_1(x_1)\psi_2(x_2)
\label{eq:TermMid1}
\end{align}
since $\comm{\hat{d}_j(y)}{\hat{b}_i^\dag(x)}=\sqrt{1-\nu}\,\delta_{ij}\delta(x-y)$.
Similarly, for $\hat{\alpha}=\hat{d}$ and $\hat{\beta}=\hat{a}$, it is non-vanishing only for $S_1=2$ and we have
\begin{align}
\mel{\mathrm{vac}}{\hat{E}^{+(\hat{d})}_{S_1C}(y\PLH)\hat{E}^{+(\hat{a})}_{\sigma(S_1)C'}(y')}{\Psi}&=\frac{\sqrt{1-\nu}}{2\pi}\int \d x_1\d x_2\d x'_1\d x'_2\  g(x_1,S_1;y\PLH,C)g(x_2,\sigma(S_1);y',C')\notag\\
& \qquad\times \psi_1(x'_1)\psi_2(x'_2)\, \delta_{2S_1}\delta(x_1-x_2')\delta(x_2-x_1')\notag\\
&=\delta_{2S_1}\frac{\sqrt{1-\nu}}{2\pi}\int \d x_1\d x_2\ g(x_1,2;y\PLH,C)g(x_2,1;y',C')\psi_1(x_2)\psi_2(x_1).
\label{eq:TermMid2}
\end{align}
Finally, for $\hat{\alpha}=\hat{a}$ and $\hat{\beta}=\hat{a}$, the term in Eq.~\eqref{eq:TermProd} becomes
\begin{align}
\mel{\mathrm{vac}}{\hat{E}^{+(\hat{a})}_{S_1C}(y\PLH)\hat{E}^{+(\hat{a})}_{\sigma(S_1)C'}(y')}{\Psi}&=\frac{1}{2\pi}\int \d x_1\d x_2\d x'_1\d x'_2\  g(x_1,S_1;y\PLH,C)g(x_2,\sigma(S_1);y',C')\psi_1(x'_1)\psi_2(x'_2)\notag\\
& \qquad\times \sqrt{\nu}\left(\delta_{1S_1}\delta(x_1-x_1')\delta(x_2-x_2')+\delta_{2S_1}\delta(x_1-x_2')\delta(x_2-x_1')\right)\notag\\
&=\frac{1}{2\pi}\Big(\sqrt{\nu}\,\delta_{1S_1}\,\int\d x_1\d x_2\ g(x_1,1;y\PLH,C)g(x_2,2;y',C')\psi_1(x_1)\psi_2(x_2)\notag\\
&\qquad+\sqrt{\nu}\,\delta_{2S_1}\,\int\d x_1\d x_2\ g(x_1,2;y\PLH,C)g(x_2,1;y',C')\psi_1(x_2)\psi_2(x_1)\Big).
\label{eq:TermMid3}
\end{align}
The integrals appearing in Eqs.~\eqref{eq:TermMid1}-\eqref{eq:TermMid3} are similar to each other, and they can be written in terms of the Fourier transforms of the position wave-functions $\psi_S(x)$.
Indeed
\begin{align}
\int \d x g(x,S;y,C)\psi_{S}(x)=\frac{1}{\sqrt{2\pi}}\frac{\e^{\ii\phi(S,C)}}{\sqrt{2}}\e^{\ii k_0 d}\int \d x\ \e^{-\ii \frac{k_0}{d} x y}\psi(x-x_{0S})=\frac{\e^{\ii\phi(S,C)}}{\sqrt{2}}\e^{\ii k_0 d}\e^{-\ii\frac{k_0}{d}y x_{0s}}\varphi\left(\frac{k_0}{d}y\right).
\label{eq:FTransf}
\end{align}
We can thus write all the second-order correlation functions in Eq.~\eqref{eq:G2Prod} by plugging in the expressions in Eq.~\eqref{eq:TermMid1}-\eqref{eq:FTransf}, obtaining 
\begingroup
\begin{align}
G^{(2)}_{CC'}(d,y\PLH;d,y')&=0\notag\\
G^{(2)}_{CC'}(a,y\PLH;d,y')&=\frac{1-\nu}{4}\abs{\varphi(k\PLH)}^2\abs{\varphi(k')}^2\notag\\
G^{(2)}_{CC'}(d,y\PLH;a,y')&=\frac{1-\nu}{4}\abs{\varphi(k\PLH)}^2\abs{\varphi(k')}^2\notag\\
G^{(2)}_{CC'}(a,y\PLH;a,y')&=\frac{\nu}{4}\abs{\varphi(k\PLH)}^2\abs{\varphi(k')}^2\big\vert\e^{\ii(\Phi(1,C)+\Phi(2,C'))}\e^{-\ii (k\PLH x_{01}+k'x_{02})}+\e^{\ii(\Phi(2,C)+\Phi(1,C'))}\e^{-\ii (k\PLH x_{02}+k'x_{01})}\big\vert^2\notag\\
&=\frac{\nu}{4}\abs{\varphi(k\PLH)}^2\abs{\varphi(k')}^2(2+2\Re(\e^{\ii(\Phi(1,C)+\Phi(2,C')-\Phi(2,C)-\Phi(1,C'))}\e^{-\ii (k\PLH-k')\Delta x})),
\label{eq:G2App1}
\end{align}
\endgroup
where we introduced the notation $k = k_0 y/d$, $k'=k_0y'/d$ and $\Delta x=x_{01}-x_{02}$.
Since the further degree of distinguishability denoted by the optical modes $\hat{a}$ and $\hat{d}$ is not resolved, we sum over all the previous terms in Eq.~\eqref{eq:G2App1} for $C=C'$ (same-camera events) and $C\neq C'$ (two-camera events).
Due to the constraint $\Phi(1,C_1)+\Phi(2,C_2)-\Phi(1,C_2)-\Phi(2,C_1)=\pm\pi$ imposed by the unitarity of the beam-splitter, we notice that the phase term $\e^{\ii(\Phi(1,C)+\Phi(2,C')-\Phi(2,C)-\Phi(1,C')}$ is always $+1$ for $C=C'$ and $-1$ for $C\neq C'$. 
We thus obtain
\begin{align}
P_\nu(k\PLH,k',A) &= \sum_{\alpha,\beta=a,d} G^{(2)}_{12}(\alpha,y\PLH;\beta,y')= \frac{1}{2}\abs{\varphi(k\PLH)}^2\abs{\varphi(k')}^2\left(1-\nu\cos((k\PLH-k')\Delta x)\right)\notag\\
P_\nu(k\PLH,k',B) &= \sum_{\alpha,\beta=a,d} \frac{G^{(2)}_{11}(\alpha,y\PLH;\beta,y')+G^{(2)}_{22}(\alpha,y\PLH;\beta,y')}{2}=\frac{1}{2}\abs{\varphi(k\PLH)}^2\abs{\varphi(k')}^2\left(1+\nu\cos((k\PLH-k')\Delta x)\right),
\label{eq:AppProbPre}
\end{align}
where the factor $1/2$ for the same-camera probabilities is needed to correctly take into account the symmetry for the exchange $k\PLH\leftrightarrow k'$, and thus to consider $P_\nu(k\PLH,k',B)$ as defined over every $k\PLH,k'$ and not, e.g., only for $k\PLH\leqslant k'$.
Notice that the probability $P(\Delta k,X)$ in Eq.~\eqref{eq:Prob} in the main text coincide with Eq.~\eqref{eq:AppProbPre} after the change of variables $K=(k\PLH+k')/2$ and $\Delta k = k\PLH-k'$, and integrating over $K$ 
\begin{multline}
P_\nu(\Delta k,X) = \frac{1}{2}\left(\int\dd K\ \abs{\varphi\left(K+\frac{\Delta k}{2}\right)}^2\abs{\varphi\left(K-\frac{\Delta k}{2}\right)}^2\right)(1+\alpha(X)\nu\cos(\Delta k\Delta x))\\
 \equiv \frac{1}{2}C(\Delta k)(1+\alpha(X)\nu\cos(\Delta k\Delta x))
\label{eq:AppProb}
\end{multline}
where, we introduced the envelope function 
\begin{equation}
C(\Delta k) = \int\dd K\ \abs{\varphi\left(K+\frac{\Delta k}{2}\right)}^2\abs{\varphi\left(K-\frac{\Delta k}{2}\right)}^2.
\label{eq:Envelope}
\end{equation}
It is easy to show that $C(\Delta k)$ reduces to
\begin{equation}
C(\Delta k) = \frac{1}{\sqrt{4\pi\sigma_k^2}}\exp(-\frac{\Delta k^2}{4\sigma_k^2})
\end{equation}
for Gaussian transverse momentum distributions $\abs{\varphi(k)}^2$.

\section{QFI in Eq.~\eqref{eq:QFI}}
\label{app:QFI}

To see that the quantum Fisher information (QFI) for the estimation of the separation $\Delta x=x_{01}-x_{02}$ of the two photons in the state in Eq.~\eqref{eq:InitState} is the one shown in Eq.~\eqref{eq:QFI} in the main text, it is sufficient to notice that the initial state in Eq.~\eqref{eq:InitState}, with $\psi_S(x)=\psi(x-x_{0S})$ is generated by a unitary translation of a state $\ket{\Psi_{0}}$, i.e.
\begin{equation}
\ket{\Psi} = \hat{U}(x_{01},x_{02})\ket{\Psi_{0}}=\hat{U}(x_{01},x_{02})\int\d x_1\ \psi(x_1)\hat{a}_1^\dag(x_1)\ket{0}_1 
\otimes \int\d x_2\ \psi(x_2)\hat{b}_2^\dag(x_2)\ket{0}_2,
\end{equation}
with
\begin{equation}
\hat{U}(x_{01},x_{02}) = \exp(-\ii x_{01} \hat{k}_1)\otimes \exp(-\ii x_{02} \hat{k}_2 )
\label{eq:Unitary1}
\end{equation}
unitary evolution generated by the conjugate variables to the photon positions, i.e., the transverse momenta
\begin{align}
\hat{k}_1\equiv\int \dd k\ k\, \hat{a}_1^\dag(k)\hat{a}_1(k)\notag\\
\hat{k}_2\equiv\int \dd k\ k\, \hat{b}_2^\dag(k)\hat{b}_2(k),
\end{align}
and $\hat{a}_1(k)$ being defined as $\int \d x\ \e^{-\ii k x} \hat{a}_1(x)$, and similarly for $\hat{b}_2(k)$.
Since we can write $x_{01}=x_{\mathrm{sum}}/2+\Delta x/2$ and $x_{02}=x_{\mathrm{sum}}/2-\Delta x/2$, with $x_{\mathrm{sum}}=(x_{01}+x_{02})$, substituting these expressions in Eq.~\eqref{eq:Unitary1}, we see that the operators
\begin{equation}
\hat{G}^\pm = \frac{\hat{k}_1}{2} \pm\frac{\hat{k}_2}{2}
\end{equation}
are the generators of the parameters $\Delta x$ and $x_{\mathrm{sum}}$, independently of $\nu$.
We will assume that no information on the parameter $x_{\mathrm{sum}}$ is available prior to the estimation of $\Delta x$, so that the quantum Cramér-Rao bound on $\Delta x$ must be evaluated through the two-parameter QFI matrix $H$~\cite{Helstrom1969, Paris2009, Holevo2011, Liu2020}.
In this type of scenario, i.e., for pure states and unitary evolutions, the $2\times 2$ QFI matrix of the parameters $\Delta x$ and $x_{\mathrm{sum}}$ is evaluated as~\cite{Liu2020}
\begin{equation}
H_{ij}=4\mathrm{Cov}_{\ket{\Psi_0}}(\hat{G}^i,\hat{G}^j),\quad i,j=\pm,
\label{eq:QFIMSupp}
\end{equation}
where $\mathrm{Cov}_{\ket{\Psi_0}}$ represents the covariance over the state $\ket{\Psi_0}$.
Since $\hat{k}_1$ and $\hat{k}_2$ commute, clearly also $\hat{G}^+$ and $\hat{G}^-$ commute. 
We thus evaluate the covariance $\mel{\Psi_0}{\hat{G}^+\hat{G}^-}{\Psi_0}-\mel{\Psi_0}{\hat{G}^+}{\Psi_0}\mel{\Psi_0}{\hat{G}^-}{\Psi_0}=0$, and the variances $\mel{\Psi_0}{\hat{G}^{\pm 2}}{\Psi_0}-\mel{\Psi_0}{\hat{G}^\pm}{\Psi_0}^2=\sigma_k^2/2$, with $\sigma_k^2$ the transverse-momentum variance of each photon.
The QFI matrix in Eq.~\eqref{eq:QFIMSupp} associated with the estimation of $x_{\mathrm{sum}}$ and $\Delta x$ is thus diagonal, meaning that these two parameters can be estimated simultaneously and independently, and in particular the QFI associated with $\Delta x$ is
\begin{equation}
H(\Delta x) = 4(\mel{\Psi_0}{(\hat{G}^{-})^2}{\Psi_0}-\mel{\Psi_0}{\hat{G}^-}{\Psi_0}^2)=2\sigma_k^2\equiv H,
\end{equation}
as shown in Eq.~\eqref{eq:QFI} in the main text.

\section{FI in Eqs.~\eqref{eq:QFI} and~\eqref{eq:FINu} and its contributions in Eq.~\eqref{eq:FIContr}}
\label{app:FI}

The Fisher information (FI) associated with the estimation of the separation $\Delta x$ through our transverse-momentum resolving technique is defined as~\cite{Cramer1999, Rohatgi2000}
\begin{equation}
F_\nu(\Delta x) = \E\left[\left(\frac{\d}{\d \Delta x}\log(P_\nu(\Delta k,X))\right)^2\right]
\equiv \sum_{X=A,B}\int \d \Delta k\ P_\nu(\Delta k,X)\left(\frac{\d}{\d \Delta x}\log(P_\nu(\Delta k,X))\right)^2,
\label{eq:FIAppDef}
\end{equation}
with $P_\nu(\Delta k,X)$ found in Eq.~\eqref{eq:AppProb}, where the case of completely indistinguishable photons is obtained by fixing $\nu=1$.
We first evaluate
\begin{equation}
\left(\frac{\d}{\d \Delta x}P_\nu(\Delta k,X)\right)^2=\frac{\nu^2}{4}C(\Delta k)^2\Delta k^2\sin^2(\Delta k\Delta x),
\label{eq:NumDerivative}
\end{equation}
where $C(\Delta k)$ takes the form found in Eq.~\eqref{eq:Envelope}.
Then, recalling the expression of the probability $P_\nu(\Delta k,X)$ in Eq.~\eqref{eq:AppProb}, we easily see that Eq.~\eqref{eq:FIAppDef} becomes, for $\nu=1$, 
\begin{equation}
F_{\nu=1}(\Delta x)=\int\dd \Delta k\ C(\Delta k)\Delta k^2\equiv\int\dd k_1\dd k_2\ \abs{\varphi(k_1)}^2\abs{\varphi(k_2)}^2(k_1-k_2)=2\sigma_k^2\equiv H,
\label{eq:FINu1}
\end{equation}
as shown in Eq.~\eqref{eq:QFI} in the main text, while, for $\nu<1$
\begin{equation}
F_\nu(\Delta x) = H \int\dd\Delta k\ C(\Delta k)\frac{\Delta k^2}{2\sigma_k^2}\frac{\nu^2\sin^2(\Delta k\Delta x)}{1-\nu^2\cos^2(\Delta k\Delta x)}\equiv H\int \dd\Delta k\ f_\nu(\Delta k,\Delta x).
\label{eq:FInuApp}
\end{equation}
thus retrieving the result in Eq.~\eqref{eq:FINu} in the main text.

\section{FI for non-resolving detectors}

If we replace the resolving cameras at the outputs of the beam splitter with simple bucket detectors, we average out all the information about the separation $\Delta x$ obtained by resolving the transverse momenta $k\PLH$ and $k'$ of the photons.
In this case, we would only discriminate two possible outcome events, i.e., photons detected by the different ($X=A$) or same ($X=B$) detector, occurring with probabilities
\begin{align}
P_\nu(X)&=\int\dd \Delta k\ P_\nu(\Delta k,X) = \frac{1}{2}\left(1+\nu\alpha(X)\int\dd \Delta k\ C(\Delta k)\cos(\Delta k\Delta x)\right)
\label{eq:NoResProb}
\end{align}
obtained integrating the probabilities in Eq.~\eqref{eq:AppProb}.
Notice that the integral in Eq.~\eqref{eq:NoResProb} defines the shape of the HOM dip, which clearly depends on the transverse-momentum distribution $\abs{\varphi(k)}^2$ of the two photons through the envelope $C(\Delta k)$.

We can now evaluate the FI associated with the estimation of $\Delta x$ through measurements that do not resolve the transverse momenta of the photons, i.e., arising from the probabilities evaluated in Eq.~\eqref{eq:NoResProb}.
In this case, the FI is a sum of only two contributions
\begin{multline}
F_\nu^{\mathrm{(nr)}}(\Delta x) = \sum_{X=A,B}P_\nu(X)\left(\frac{\dd}{\dd\Delta x}\log P_\nu(X)\right)^2=\nu^2\sum_{X=A,B} \frac{1}{2}\frac{\left(\int\dd \Delta k\ C(\Delta k)\Delta k\sin(\Delta k\Delta x)\right)^2}{1+\alpha(X)\nu\int\dd \Delta k\ C(\Delta k)\cos(\Delta k\Delta x)}\\
=\nu^2\frac{\left(\int\dd \Delta k\ C(\Delta k)\Delta k\sin(\Delta k\Delta x)\right)^2}{1-\nu^2\left(\int\dd \Delta k\ C(\Delta k)\cos(\Delta k\Delta x)\right)^2},
\label{eq:NRFI}
\end{multline}
and it depends on the transverse momentum distribution $\abs{\varphi(k)}^2$ through the envelope $C(\Delta k)$.

In the regime of small separations $\Delta x\sigma_k\ll 1$ such that the two spatial wavepackets are mostly overlapping, and $\nu=1$, we can neglect higher terms of $\Delta x$ in Eq.~\eqref{eq:NRFI}, obtaining
\begin{equation}
F_{\nu=1}^{\mathrm{(nr)}}(\Delta x)\simeq\frac{\Delta x^2\left(\int\dd \Delta k\ C(\Delta k)\Delta k^2\right)^2}{\Delta x^2\int \d \Delta k\ C(\Delta k)\Delta k^2}=2\sigma_k^2\equiv H,
\end{equation}
which is equal to the QFI in Eq.~\eqref{eq:QFI} in the main text.
Therefore, for small separations, i.e. such that $\sigma_k\Delta x\ll 1$, it is even possible, for approximately identical photons, to replace the resolving cameras with bucket detectors without any loss of sensitivity.

\section{FI for spatially resolved measurements in the regime $\sigma_k\Delta x\gg 1$} 

Here, we will evaluate the asymptotic expressions of $F_\nu(\Delta x)$ in Eq.~\eqref{eq:FINu} in the regime of large separations $\Delta x$.
For $\nu=1$, the FI in~\eqref{eq:FINu1} is constantly equal to $F_{\nu=1}(\Delta x)=2\sigma_k^2\equiv H$.
For $\nu<1$, Eq.~\eqref{eq:FInuApp} can be written as
\begin{align}
\frac{F_\nu(\Delta x)}{H} =\int\dd\Delta k\ C(\Delta k)\frac{\Delta k^2}{2\sigma_k^2}\beta_\nu(\Delta k;\Delta x)
\label{eq:FIABContrApp}
\end{align}
where
\begin{equation}
\beta_\nu(\Delta k;\Delta x)=
\dfrac{\nu^2\sin^2(\Delta x\Delta k)}{1-\nu^2\cos^2(\Delta x\Delta k)}
\label{eq:GammaDef}
\end{equation}
is a periodic function in $\Delta k$ of period inversely proportional to $\Delta x$, i.e. $\pi/\Delta x$.
We can thus split the interval of integration in Eq.~\eqref{eq:FIABContrApp} in sub-intervals of width $\pi/\Delta x$, so that the overall integral can be written as sum
\begin{equation}
\frac{F_\nu(\Delta x)}{H}=\sum_{n\in\mathbb{Z}}\int_{\frac{n\pi}{\Delta x}}^{\frac{(n+1)\pi}{\Delta x}} \d \Delta k\ C(\Delta k)\frac{\Delta k^2}{2\sigma_k^2}\beta_\nu(\Delta k;\Delta x),
\end{equation}
with $n$ integer.
If the transverse-momentum distribution $\abs{\varphi(k)}^2$ does not have `atoms', i.e. it has no discrete part, and thus the same is true for $C(\Delta k)$, $C(\Delta k)\Delta k^2$ can be eventually be considered constant within each integration domain $[n\pi/\Delta x,(n+1)\pi/\Delta x]$ for large $\Delta x$.
We can thus approximate $\Delta k^2 C(\Delta k)\simeq(\pi n/\Delta x)^2 C(\pi n/\Delta x)$ for each integral, so that
\begin{equation}
\frac{F_\nu(\Delta x)}{H} \simeq \frac{1}{2\sigma_k^2}\sum_{n\in\mathbb{Z}}\left(\frac{\pi n}{\Delta x}\right)^2 C\left(\frac{\pi n}{\Delta x}\right)\int_{\frac{n\pi}{\Delta x}}^{\frac{(n+1)\pi}{\Delta x}} \d \Delta k\ \beta_\nu(\Delta k;\Delta x,X)
=\frac{1}{2\sigma_k^2}\sum_{n\in\mathbb{Z}}\left(\frac{\pi n}{\Delta x}\right)^2 C\left(\frac{\pi n}{\Delta x}\right) \frac{\pi}{\Delta x}\bar{\beta}_\nu,
\label{eq:LargeSepStep}
\end{equation}
where, employing Eq.~\eqref{eq:GammaDef},
\begin{equation}
\bar{\beta}_\nu=1-\sqrt{1-\nu^2}
\label{eq:GammaNu}
\end{equation}
is the average of $\beta_\nu(\Delta k;\Delta x)$ over its period.
We can now rewrite the summation in Eq.~\eqref{eq:LargeSepStep} as an integral with the substitutions $\pi n/\Delta x \rightarrow \zeta$, $\sum_{n\in\mathbb{Z}}\pi /\Delta x \rightarrow \int\d \zeta$ yielding
\begin{equation}
\frac{F_\nu(\Delta x)}{H}\simeq \bar{\beta}_\nu\frac{1}{2\sigma_k^2}\int\d \zeta\ C(\zeta)\zeta^2=\bar{\beta}_\nu.
\label{eq:LargeAsym}
\end{equation}

\subsection{Contributions to the FI from the same or different cameras}
\label{app:ReducedFI}

In this section we will show how the FI is affected if, in a given experiment, some possible events are undetected or neglected.
We will use the results of this section to show that single-camera detection events are more informative than two-camera detection events, and thus to obtain the FI for the one-camera variant of our setup shown in the main text.

Let us imagine an experiment whose outcomes follow a discrete probability, and that we arbitrarily partition the outcomes in two groups.
The case of a continuous probability distribution can be obtained with a simple change of notation.
Each outcome has a given probability to happen that depend on an unknown parameter $\lambda$ that we want to estimate.
Thus, the probability of a given outcome $i$ of the first group is given by $P_{1,i}(\lambda)$, while the probability of an outcome $i'$ of the second group is $P_{2,i'}(\lambda)$, with $i=1,\dots,n$, $i'=1,\dots,n'$, and $\sum_{i=1}^n P_{1,i}(\lambda)+\sum_{i'=1}^{n'} P_{2,i'}(\lambda)=1$.
For simplicity, from now on, we will omit to explicit the dependency from $\lambda$ of the probabilities $P_{1,i}(\lambda)\equiv P_{1,i}$, $P_{2,i'}(\lambda)\equiv P_{2,i'}$. 

If none of the possible outcomes are neglected, and the information on $\lambda$ is retrieved from the outcomes of both groups, the FI associated with the estimation of $\lambda$ after $N$ repetitions of the experiment is, by definition,
\begin{equation}
F(\lambda) = N\left(\sum_{i=1}^n \frac{(\frac{\d}{\d \lambda}P_{1,i})^2}{P_{1,i}}+ \sum_{i'=1}^{n'} \frac{(\frac{\d}{\d \lambda}P_{2,i'})^2}{P_{2,i'}}\right).
\label{eq:FisherGlobal}
\end{equation}
We want now to compare $F(\lambda)$ with the FI $F_1(\lambda)$ associated with the observation of only the outcomes of the first group.
If we are neglecting the outcomes of the second group, only $N\sum_{i=1}^n P_{1,i}$ outcomes will be observed in average, while the (conditioned) probability for the $i$-th event to happen is given by $P_{1,i}/\sum_{j=1}^n P_{1,j}$.
We can then evaluate $F_1(\lambda)$ as
\begin{align}
F_1(\lambda) &= \left(N\sum_{k=1}^n P_{1,k}\right)\left(\sum_{i=1}^n\frac{P_{1,i}}{\sum_{j=1}^n P_{1,j}}\left(\frac{\d}{\d\lambda}\log(\frac{P_{1,i}}{\sum_{j=1}^n P_{1,j}})\right)^2\right)\notag\\
&=N \left(\sum_{i=1}^nP_{1,i}\left(\frac{\d}{\d\lambda}\log(P_{1,i})-\frac{\d}{\d\lambda}\log(\sum_{j=1}^n P_{1,j})\right)^2\right)\notag\\
&=N\left(\sum_{i=1}^n \frac{(\frac{\d}{\d \lambda}P_{1,i})^2}{P_{1,i}} - \frac{\left(\frac{\d}{\d\lambda}\sum_{i=1}^n P_{1,i}\right)^2}{\sum_{i=1}^n P_{1,i}}\right)\notag\\
&\equiv N\left(\sum_{i=1}^n \frac{(\frac{\d}{\d \lambda}P_{1,i})^2}{P_{1,i}} - \frac{\left(\frac{\d}{\d\lambda}P_1\right)^2}{P_1}\right)\leqslant F(\lambda)
\label{eq:FIGroups}
\end{align}
where we introduced $P_1=\sum_{i=1}^n P_{1,i}$ the overall probability that any outcome of the first group is observed.
We recognise in Eq.~\eqref{eq:FIGroups} the first term of the total FI in Eq.~\eqref{eq:FisherGlobal}, given by the sum of the contributions of each outcome of the first group.
However, this is in general reduced by a term $N\left(\frac{\d}{\d\lambda}P_1\right)^2/P_1$ that can be interpreted as the amount of information that is lost due to the fact that we are neglecting all the events of the second group, and thus we do not gain any information on the value of $P_1$.
However, if $P_1$ does not depend on $\lambda$, or in general has a negligible derivative, this term can be neglected and the FI $F_1(\lambda)$ is equal to the sum of the contributions of the outcomes of the first group.

This result can be applied to our scheme. 
Once we identify $P_{1,i} \equiv P_\nu(\Delta k,X)$, i.e. we assume that we only observe one-camera detection events ($X=B$) or two-camera detection events ($X=A$), we can apply Eq.~\eqref{eq:FIGroups} to obtain the corresponding FI, plotted in \figurename~\ref{fig:FPartial} in the main text
\begin{equation}
F_\nu(X;\Delta x) = \int \dd \Delta k\ P_\nu(\Delta k,X)\left(\frac{\dd}{\dd \Delta x} \log P_\nu(\Delta k,X)\right)^2-P_\nu(X)\left(\frac{\dd}{\dd \Delta x}\log P_\nu(X)\right)^2\leqslant F_\nu(\Delta x),
\label{eq:FIPartialApp}
\end{equation}
where $P_\nu(X)$ is the total probability to observe two photons in the same camera ($X=B$) or different cameras ($X=A$) shown in Eq.~\eqref{eq:NoResProb}.

Finally, starting from Eq.~\eqref{eq:FIPartialApp}, we can evaluate the FI in the main text associated with the variant of our scheme which only employs a single camera at the output of the beam splitter.
In this scenario, due to the symmetrical nature of the interference, the probability $P^{\mathrm{SC}}_\nu(\Delta k)$ to observe the two photons on the only camera, with a difference of transverse momenta $\Delta k$, is exactly equal to $P^{\mathrm{SC}}_\nu(\Delta k)=P_\nu(\Delta k,B)/2$.
It is thus straightforward to see, employing Eq.~\eqref{eq:FIPartialApp}, that the FI associated with such a probability is $F(B;\Delta x)/2$, as discussed in the main text.

\section{Maximum-likelihood estimation}

In this section we will showcase the maximum-likelihood approach we undertake in order to perform numerical simulations of our estimation scheme, and thus obtain \figurename~\ref{fig:Saturation} in the main text.

Let us assume to collect statistic for $N$ iterations of the experiment, i.e. employing $N$ pairs of photons.
We denominate the $i$-th observed sample event $s_i=(\Delta k_i,\mathrm{X}_i)$, with $i=1,\dots,N$, which contains the information regarding the difference of transverse momenta $\Delta k_i$ of the two detected photons, and whether they ended up in the same output channel of the beam-splitter or not, i.e. $\mathrm{X}_i = \mathrm{A}$ for same-camera or $\mathrm{X}_i = \mathrm{B}$ for two-camera detection events.
The probability of observing the sample $s_i$ according to the probability distribution given by Eq.~\eqref{eq:Prob} in the main text, is given by
\begin{equation}
P_\nu(s_i,\Delta x)= P_\nu(\Delta k_i,X_i).
\end{equation}

Since each repetition of the experiment is performed independently of the others, the probability $P_\nu(S_{N},\Delta x)\delta k^{N}$ associated with the set $S_N=\{s_i\}_{i=1,\dots,N}$ of $N$ samples is given by the product 
\begin{equation}
P_\nu(S_N,\Delta x) = \prod_{i=1}^{N}P_\nu(s_i,\Delta x)\equiv \mathcal{L}_\nu(\Delta x|S_{N})
\label{eq:ProbabilitySample}
\end{equation}
of the single event probabilities.
The expression in Eq.~\eqref{eq:ProbabilitySample}, as a function of $\Delta x$, is called likelihood function.
Noticeably, for every possible value $\Delta x$, the likelihood $\mathcal{L}_\nu(\Delta x| S_{N})$ is a number which can be easily evaluated through Eq.~\eqref{eq:ProbabilitySample} once the data sample $S_{N}$ has been experimentally obtained.

The maximum-likelihood estimator $\widetilde{\Delta x}$ thus consists in retrieving the value of $\Delta x$ that maximises the likelihood function, i.e. it is implicitly defined by
\begin{equation}
\mathcal{L}_\nu(\widetilde{\Delta x}|S_{N}) = \sup_{\Delta x}\mathcal{L}_\nu(\Delta x|S_{N}).
\label{eq:LikelihoodEquation}
\end{equation}
The advantage of employing the maximum-likelihood estimator is that for large numbers of samples, i.e. large $N$, its expectation value equals the true value of the unknown separation $\Delta x$, and its variance saturates the Cramér–Rao Bound, i.e. $\mathrm{E}[\widetilde{\Delta x}] = \Delta x$ and $\Var{\widetilde{\Delta x}}=(N F_\nu(\Delta x))^{-1}$.
On the other hand, Eq.~\eqref{eq:LikelihoodEquation} cannot be solved analytically in general, e.g. in the case of Gaussian spectra we examined previously, so a numerical approach is usually undertaken to maximise the likelihood function $\mathcal{L}_\nu(\Delta x|S_{N})$, leading to the results described in \figurename~\ref{fig:Saturation} in the main text.

\end{document}